\documentclass[aip,jcp, numerical, preprint]{revtex4-1}
\usepackage{graphicx}
\graphicspath{{images/}}
\usepackage[table,xcdraw]{xcolor}
\usepackage{amsmath,amsfonts}
\usepackage{mathrsfs}
\usepackage[version=3]{mhchem}
\usepackage{mathptmx, bm}
\usepackage{float}
\usepackage{enumerate}
\usepackage{enumitem}
\usepackage{booktabs}

\usepackage{pifont}% http://ctan.org/pkg/pifont
\begin{document}
\newcommand \e[1]{{\textbf{#1}}}
\newcommand \bo[1]{{\bf{{#1}}}}
\newcommand \bra[1] {\left\langle {#1} \right\vert}
\newcommand \ket[1] {\left\vert {#1} \right\rangle }
\newcommand \braket[2] {\left\langle {#1} \vert{#2} \right\rangle}
\newcommand{\RNum}[1]{\uppercase\expandafter{\romannumeral #1\relax}}

%\preprint{}

\title{The Spin-MInt Algorithm: an Accurate and Symplectic Propagator for the Spin-Mapping Representation of Nonadiabatic Dynamics}
% Force line breaks with \\
\author{Lauren E. Cook}
\author{James R. Rampton}
\author{Timothy J. H. Hele}
 \email{\href{mailto:t.hele@ucl.ac.uk}{t.hele@ucl.ac.uk}}
 \affiliation{Department of Chemistry, University College London, Christopher Ingold Building, London WC1H 0AJ, United Kingdom}

\date{\today}

\begin{abstract}
Mapping methods, including the Meyer--Miller--Stock--Thoss (MMST) mapping and spin-mapping, are commonly utilised to simulate nonadiabatic dynamics by propagating classical mapping variable trajectories. Recent work confirmed the Momentum Integral (MInt) algorithm is the only known \emph{symplectic} algorithm for the MMST Hamiltonian. To our knowledge, no symplectic algorithm has been published for the spin-mapping representation without obtaining Cartesian variables and utilising the MInt algorithm. 
Here, we present the Spin-MInt algorithm which directly propagates the spin-mapping variables. First, we consider a two-level system which maps onto a spin-vector on a Bloch sphere. Despite the spin-variables being non-canonical, we rigorously prove the Spin-MInt is \emph{symplectic} using a canonical variable transformation. We determine that the Spin-MInt is a \emph{symmetrical, second-order, time-reversible, angle invariant and geometric structure preserving} algorithm. Computationally, for a one-dimensional spin-boson model, the Spin-MInt and MInt algorithms are symplectic, satisfy Liouville’s theorem, provide second-order energy conservation and are more accurate than a previously-published angle-based algorithm. We present accurate correlation functions for a multi-dimensional spin-boson model. We also extend this methodology to a general number of electronic states and present accurate population results for a three-state Morse potential. The Spin-MInt is faster than the MInt algorithm for all tested models, particularly so for large nuclear degrees of freedom.
We believe this to be the \emph{first known symplectic} algorithm for propagating the nonadiabatic spin-mapping Hamiltonian and one of the first rigorously symplectic algorithms in the case of non-trivial coupling between canonical and spin systems. These results should guide and improve future simulations.
\end{abstract}


\maketitle %\maketitle must follow title, authors, abstract and \pacs

\allowdisplaybreaks

\section{Introduction}
Nonadiabatic dynamics governs many interesting phenomena involving light and energy transfer. Simulation is particularly challenging to implement due to the coupled nuclear and electronic degrees of freedom (DoF). Full quantum methods, such as Multi-Configurational Time-Dependent Hartree Fock (MCTDH) and wavepacket methods,\cite{Beck2000,vanhaeftenPropagatingMultidimensionalDensity2023,wangMultilayerFormulationMulticonfiguration2003, shinMultipleTimeScale1996, vanhaeftenPropagatingMultidimensionalDensity2023, mukherjeeAssessingNonadiabaticDynamics2025} are extremely accurate but limited to smaller systems by the large computational expense associated. For single surface (adiabatic) systems, many classical-scaling methods have been introduced including ring-polymer molecular dynamics (RPMD),\cite{Craig2004, Habershon2013, Hele2015, heleAlternativeDerivationRingpolymer2016} the associated centroid molecular dynamics (CMD),\cite{ Cao1993, Cao1994, Cao1994a, Cao1994b,heleCommunicationRelationCentroid2015, jangPathIntegralCentroid1999, jangDerivationCentroidMolecular1999, Jung2020, castro_vibrational_2025} and thermostatted (T)-RPMD,\cite{Rossi2014, Hele2015c, Hele2016a, ceriottiEfficientStochasticThermostatting2010} and the Linearised Semiclassical Initial-Value Representations (LSC-IVR).\cite{ Stock1997, Miller1970,Sun1998, Sun1998b, Wang1998, Wang1999} Several extensions for nonadiabatic dynamics are available such as mean-field RPMD, nonadiabatic (N)-RPMD and mapping variable (MV)-RPMD for RPMD,\cite{Hele2011, Richardson2013, Richardson2017, Ananth2013} and Mixed Quantum Classical (MCQ)-IVR for IVR methods.\cite{Ananth2007, Church2017, Antipov2015, Liu2015, Filinov1986, Thoss2001, Shi2004} Additionally, several mixed quantum-classical methods exist including Ehrenfest methods, surface hopping and mapping methods.\cite{Kapral2016, Tully1971, tullyPerspectiveNonadiabaticDynamics2012, Shakib2017, Shalashilin2011, Zimmermann2014, Meyer1979,Stock1997, Stock2005, Runeson2019, MASH, althorpeNonadiabaticReactionsGeneral2017, Althorpe2016,nelsonNonadiabaticExcitedStateMolecular2020} 

Here, we focus on trajectory-based mapping methods which map the discrete quantum system onto continuous variables which can be propagated classically. One well-established method is the Meyer-Miller-Stock-Thoss (MMST) mapping which obtains electronic positions and momenta and evolves under the MMST Hamiltonian using Hamilton's equations of motion (EOM).\cite{Meyer1979, Stock1997} Alternatively, spin-mapping obtains a spin-vector on the surface of a Bloch sphere (in the two-level case) which follows Heisenberg's EOM.\cite{Runeson2019, Runeson2020, Runeson2021} This is an advantageous representation as it removes two redundant DoF compared to the MMST representation.\cite{Runeson2019, Runeson2020} The two representations are related, where elements of the spin-vector can be calculated from MMST variables and the Hamiltonians are equivalent under the addition of a zero-point energy (ZPE) parameter to the MMST Hamiltonian.\cite{Runeson2019, Runeson2020, bossionNonadiabaticMappingDynamics2022} A ZPE fitting parameter was first introduced by Stock and M\"uller to mitigate ZPE leakage, when energy unphysically flows between states, and many possibilities were tested.\cite{Muller1999, Stock1999} However, spin-mapping derives this as a function of the number of states and leads to a parameter that is close to what was previously found to be optimal.\cite{Runeson2019}

Historically, spin-mapping was limited to two electronic states.\cite{Runeson2019, Bossion2021} However, spin-coherent states and the $SU(N)$ Lie group have been harnessed to generalise to more states and we use this approach to extend this work.\cite{Runeson2020, bossionNonadiabaticMappingDynamics2022, bossionNonadiabaticRingPolymer2023} While many of the nonadiabatic ring-polymer methods utilise the MMST mapping, a spin-mapping (SM)-NRPMD has been proposed,\cite{Bossion2021, bossionNonadiabaticRingPolymer2023} which alongside the flurry of spin-mapping methods including a partially linearized approach,\cite{Mannouch2020, Mannouch2020b, Mannouch2022} an ellipsoidal method,\cite{Amati2023} generalisation to multiple states,\cite{bossionNonadiabaticMappingDynamics2022, bossionNonadiabaticRingPolymer2023, Runeson2020} and the Mapping Approach to Surface Hopping (MASH), highlights the interest in spin-mapping approaches.\cite{MASH, lawrenceSizeconsistentMultistateMapping2024, geutherTimeReversibleImplementationMASH2025, richardsonNonadiabaticDynamicsMapping2025, runesonExcitonDynamicsMapping2024, furlanettoSimulatingElectronicCoherences2025} As far as we are aware, none of the mapping methods developed thus far reproduce Rabi oscillations, scale classically with system size, conserve the quantum Boltzmann distribution, have a clear derivation from exact quantum dynamics and produce accurate dynamics.\cite{cookElectronicPathIntegral2025} Recent work showed that such a method is unlikely to be obtained through use of MMST variables and the related electronic state descriptors but an alternative metric, such as a spin-mapping metric, may achieve this in future.\cite{cookElectronicPathIntegral2025}

To investigate dynamical properties of nonadiabatic systems, mapping methods are often utilised to approximate correlation functions of the form, 
\begin{align}
    \label{General-CF}
    C_{AB}= \frac{1}{Z} \mathrm{Tr}[e^{-\beta \hat{H}}\hat{A}(0)\hat{B}(t)]\text{,}
\end{align}
where $e^{-\beta \hat{H}}$ is the quantum Boltzmann distribution at inverse temperature, $\beta = 1/ k_B T$, $\hat{A}$ and $\hat{B}$ are operators and $Z$ is the partition function given by,
\begin{align}
    \label{partiton}
    Z = \mathrm{Tr} [e^{-\beta \hat{H}}] \text{.}
\end{align}
To approximate these, many trajectories are required to ensure that enough phase-space is sampled. Therefore, we need accurate propagation algorithms.\cite{Cook2023} 

For Hamiltonian systems, algorithms can be \emph{symplectic} which arises from exact Hamiltonian or sub-Hamiltonian integration and results in little/no energy drift over time when propagating the dynamics.\cite{Leimkuhler2005} This is also advantageous for methods that have a phase-factor as a symplectic integrator aids convergence.\cite{Church2018, Bonella2001, Miller1970} Although many algorithms have been developed for the MMST Hamiltonian,\cite{Church2018, Kelly2012, Richardson2017, Wang1999} a recent study has shown that the Momentum Integral (MInt) algorithm is the only rigorously symplectic integrator.\cite{Church2018, Cook2023} As far as we are aware, no such algorithm has been proposed for the spin-mapping representation.\cite{Runeson2019} While an angle-based algorithm has been utilised for the spin-mapping Hamiltonian, it is known to have instabilities and it was subsequently suggested that to obtain symplectic propagation, one should sample MMST variables from the hypersphere and propagate with the MInt algorithm.\cite{bossionNonadiabaticMappingDynamics2022, Bossion2021, Runeson2020} We believe it would be advantageous to remain in the spin-mapping representation by direct propagation of the spin-variables rather than introduce redundant DoF. Other algorithms have been suggested for general spin-systems,\cite{tranchidaMassivelyParallelSymplectic2018, prasadNewSymplecticIntegrator2023, Ivanov2013, lubichSymplecticIntegrationPostNewtonian2010, mclachlanSymplecticIntegratorsSpin2014, frank_geometric_1997, huangSymplecticSpinLatticeDynamics2025} but many assume a separable Hamiltonian, use a Split-Liouvillian approximation (which we later show results in a loss of symplecticity for the spin-mapping Hamiltonian) or are not applicable to a system with a non-trivial coupling between the nuclear and electronic DoF. Hence, we do not consider them further in this work.

We motivate our work through the following. Prior to this work the only published algorithm for spin-mapping trajectories (the angle-based algorithm) is known to be unstable, hence, an accurate algorithm is desirable. To obtain a symplectic algorithm, the current best practice is to sample Cartesian variables from a sphere and utilise the MInt algorithm. The spin-mapping observables are then calculated in terms of these Cartesian variables, inherently encoding the ZPE parameter.\cite{Runeson2019, bossionNonadiabaticMappingDynamics2022} In addition, as far as we are aware there is no clear literature on how to obtain symplectic propagation of a spin-system when it is non-trivially coupled to a canonical system. In this article, we present a spin-mapping approach to the MInt algorithm (the Spin-MInt algorithm) which we believe is the first symplectic algorithm for the spin-mapping Hamiltonian, following a similar flowmap to the MInt algorithm and inheriting many of its properties. The advantages of this method are that it is more accurate than the previously published angle-based algorithm, one can directly propagate the spin-mapping variables and, to the best of our knowledge, it is the fastest known algorithm for symplectic propagation of a spin system non-trivially coupled to a Cartesian system, with the speed-up being particularly pronounced for large nuclear DoF. 

The structure of this article is as follows. In section \ref{backgroundtheory}, we present background theory, including the MInt algorithm and spin-mapping. In section \ref{methodology}, we present the Spin-MInt algorithm (generalised to $N$ states in Appendix~\ref{multiple spin mint}) and its properties. Algebraically, we determine that the Spin-MInt is symplectic via a canonical variable transformation.\cite{bossionNonadiabaticMappingDynamics2022} The models used are presented in section \ref{models}. In section \ref{results}, we present computational results for two and three-level systems, finding that the Spin-MInt is faster than the MInt for all tested models, particularly for large nuclear DoF. We conclude in section \ref{conclusions}.  

\section{Background Theory} \label{backgroundtheory}
Here, we present the background theory of the MMST mapping, the MInt algorithm and spin-mapping. For simplicity, we present the two-state case of spin-mapping and discuss generalisation in Appendix~\ref{multiple spin mint}.

% As spin-mapping was first introduced for a two-state system,\cite{Runeson2019} resulting in a three dimensional spin-vector, we present this section with two electronic DoF. 
% This also allows easy computational comparison between algorithms on the simplest system.\cite{Cook2023} 

\subsection{MMST Representation}
\subsubsection{MMST Hamiltonian} \label{Hamiltonian}

The MMST Hamiltonian is
\begin{align} \label{full-MMST}
    H_{\textrm{MMST}} = \frac{1}{2} \bo{P}^\textrm{T}\boldsymbol{\mu}^{-1}\bo{P} + V_{0}(\bo{R}) + \frac{1}{2}\left\{ \bo{p}^\textrm{T}\bo{V}(\bo{R})\bo{p} + \bo{q}^\textrm{T}\bo{V}(\bo{R})\bo{q} - \gamma \textrm{Tr}[\bo{V}(\bo{R})]\right\} \text{,}
\end{align}
where $\bo{R}$ and $\bo{P}$ are $F$ dimensional vectors of nuclear position and momenta and the mapping variables, $\bo{q}$ and $\bo{p}$, are $N$ dimensional vectors of electronic position and momenta. Hence, $\boldsymbol{\mu}$ is a $F \times F$ diagonal matrix of nuclear masses. The ZPE parameter, $\gamma$, is unity in the standard MMST mapping.\cite{Muller1999} The diabatic potentials, $V_{0}(\bo{R})$ and $\bo{V}(\bo{R})$, are dependent on nuclear position and are the state-independent and state-dependent potentials respectively. For a two-level system, $\bo{V}(\bo{R})$ typically has the following form, 
\begin{align} \label{diabatic pot mat}
    \bo{V}(\bo{R}) = \begin{bmatrix}
        V_1(\bo{R}) & \Delta^*(\bo{R}) \\
        \Delta(\bo{R}) & V_2(\bo{R})
    \end{bmatrix} \text{,}
\end{align}
where $\Delta$ is the coupling between electronic states, $^*$ represents the complex conjugate and $V_{1/2}$ are the state energies.

\subsubsection{MInt Algorithm} \label{MInt}
The MInt algorithm is the only known symplectic algorithm for propagation of the MMST Hamiltonian.\cite{Church2018, Cook2023}  The Hamiltonian is split into two parts such that, 
\begin{subequations} \label{split-ham-mmst}
\begin{align}
    H_{1,\textrm{MMST}} &= \frac{1}{2} \bo{P}^\textrm{T}\boldsymbol{\mu}^{-1}\bo{P} \text{,} \\
    H_{2,\textrm{MMST}} &= V_{0}(\bo{R}) + \frac{1}{2}\left\{ \bo{p}^\textrm{T}\bo{V}(\bo{R})\bo{p} + \bo{q}^\textrm{T}\bo{V}(\bo{R})\bo{q} - \gamma \textrm{Tr}[\bo{V}(\bo{R})]\right\} \text{,}
\end{align}
\end{subequations}
and the MInt algorithm integrates both sub-Hamiltonians exactly. Here, we include the ZPE parameter for direct comparison with the Spin-MInt algorithm. If the diabatic potential matrix is traceless, which is common in spin-mapping approaches,\cite{Runeson2019} the last term is zero. The flow map for the algorithm is as follows,\cite{Church2018, Cook2023}
\begin{align}
    \label{flow-map-mint}
    \Psi^{\textrm{MInt}}_{H, \Delta t } := \Phi_{H_{1,\textrm{MMST}}, \Delta t/2} \circ \Phi_{H_{2,\textrm{MMST}}, \Delta t} \circ \Phi_{H_{1,\textrm{MMST}}, \Delta t/2} \text{,}
\end{align}
where the overall evolution for a timestep, $\Delta t$, is performed by sandwiching the evolution of $H_{2,\textrm{MMST}}$ with two half timestep evolutions of $H_{1,\textrm{MMST}}$. In this notation $\Phi$ represents exact sub-evolution and $\Psi$ is approximate evolution.\cite{Cook2023} Evolution can be performed with the two sub-Hamiltonians swapped at an increased computational cost, as shown in the Supplementary Material Table S.6.\cite{Church2018, Cook2023} 

The propagation equations are based on Hamilton's EOM where,\cite{Stock2005, Stock1997, Cook2023}
\begin{subequations} \label{Hamilton's eom}
\begin{align}
    \dot{\bo{R}} = \frac{\partial H}{\partial \bo{P}} \quad &\text{,} \quad \dot{\bo{P}} = -\frac{\partial H}{\partial \bo{R}} \text{,} \\
    \dot{\bo{q}} = \frac{\partial H}{\partial \bo{p}} \quad &\text{,} \quad  \dot{\bo{p}} = - \frac{\partial H}{\partial \bo{q}} \text{,}
\end{align}
\end{subequations}
which can be integrated exactly for each sub-Hamiltonian.  

For $H_{1,\textrm{MMST}}$, the only propagated variable is the nuclear position where, 
\begin{align}
    \label{R-prop-h1}
    R_k(t + \Delta t/2) = R_k(t) + \frac{P_k}{2\mu_{kk}}\Delta t \text{,}
\end{align}
and all other variables are unchanged.\cite{Church2018, Cook2023} 

For $H_{2,\textrm{MMST}}$, the nuclear position is unchanged and all other variables are propagated using, 
\begin{subequations} \label{H2-prop}
    \begin{align}
        \dot{P}_k &= -V^k_{0} - \frac{1}{2}\left\{  (\bo{q} - i\bo{p})^\textrm{T}\bo{V}^k(\bo{q} + i\bo{p})  - \gamma \textrm{Tr}[\bo{V}^k]\right\} \text{,} \\ \label{H2-ELEC}
        \dot{\bo{q}} &= \bo{V}\bo{p} \quad \text{,} \quad \dot{\bo{p}} = -\bo{V}\bo{q} \text{,}
    \end{align}
\end{subequations}
where $\bo{V}^k$ is the derivative with respect to $R_k$ and likewise for any superscript of $k$. Note that we drop the dependence on $\bo{R}$ for simplicity from here. The MInt algorithm utilises the fact that $\dot{\bo{q}}$ and $\dot{\bo{p}}$ are independent of $\bo{P}$ but $\dot{\bo{P}}$ depends on $\bo{q}$ and $\bo{p}$ such that one can solve for time-evolved $\bo{q}$ and $\bo{p}$ and then use these values to solve for time-evolved $\bo{P}$.\cite{Cook2023, Church2018} 

The electronic propagation in Eqn.~\eqref{H2-ELEC} becomes, 
\begin{align} \label{q_p_prop}
    [\bo{q} + i\bo{p}](t + \Delta t) = e^{-i\bo{V}\Delta t} [\bo{q} + i \bo{p}](t) \text{,}
\end{align}
as is seen for many MMST integrators,\cite{Richardson2013, Richardson2017, Kelly2012, Church2018, Cook2023} and we use the shorthand $[\bo{q} + i\bo{p}](t) \equiv \bo{q}(t) + i\bo{p}(t)$.  The decomposition of the diabatic potential matrix is,
\begin{align} \label{decomposition}
    \bo{V} = \bo{S} \boldsymbol{\Lambda} \bo{S}^\mathrm{T} \text{,}
\end{align}
where $\bo{S}$ is the eigenvectors and $\boldsymbol{\Lambda}$ is a diagonal matrix of the eigenvalues. In Section IA of the Supplementary Material, an algebraic form of $\bo{S}$ and $\boldsymbol{\Lambda}$ and their derivatives for the potential matrix in Eqn.~\eqref{diabatic pot mat} where $\Delta$ is real is presented. Hence,
\begin{subequations} \label{q_p_mint}
    \begin{align}
        \bo{q}(t+ \Delta t ) &= \bo{C}\bo{q}(t) - \bo{D}\bo{p}(t) \text{,} \\
        \bo{p}(t+ \Delta t ) &= \bo{C}\bo{p}(t) + \bo{D}\bo{q}(t) \text{,}
    \end{align}
\end{subequations}
where, 
    \begin{align} \label{c_d}
        \bo{C} = \bo{S}\cos (\boldsymbol{\Lambda}\Delta t)\bo{S}^{\textrm{T}} \quad \text{,} \quad
        \bo{D} = -\bo{S}\sin (\boldsymbol{\Lambda}\Delta t)\bo{S}^{\textrm{T}} \text{.}
    \end{align}
For the nuclear momentum propagation, Eqn.~\eqref{q_p_prop} is used to obtain, 
 \begin{align} \label{p_int}
        P_k(\Delta t) &= -\left\{V^k_{0} - \frac{\gamma}{2} \textrm{Tr}[\bo{V}^k]\right\} \Delta t - \frac{1}{2} \int_0^{\Delta t} \textrm{d}t \left\{  [\bo{q} - i\bo{p}]^\textrm{T}(0) e^{i\bo{V} t} \bo{V}^k e^{-i\bo{V} t}[\bo{q} + i\bo{p}](0)   \right\} \text{,}
    \end{align}
The derivative of the potential in the adiabatic basis is defined as, $\bo{G}^k = \bo{S}^\mathrm{T} \bo{V}^k \bo{S}$, and inserting $\bo{S}\bo{S}^\mathrm{T}$ identities, 
\begin{align}
    P_k(\Delta t) 
    =& P_k(0) -\left\{V^k_{0} - \frac{\gamma}{2} \textrm{Tr}[\bo{V}^k]\right\} \Delta t \nonumber \\ &- \frac{1}{2} \int_0^{\Delta t} \textrm{d}t \left\{  [\bo{q} - i\bo{p}]^\textrm{T}(0) \bo{S}e^{i\boldsymbol{\Lambda}t}\bo{G}^ke^{-i\boldsymbol{\Lambda}t}\bo{S}^\mathrm{T}[\bo{q} + i\bo{p}](0)   \right\} \text{,}
\end{align}
where the MInt approach to the integral is,
\begin{align}
    I = \int_0^{\Delta t} \mathrm{d}t \quad \bo{S}e^{i\boldsymbol{\Lambda}t}\bo{G}^ke^{-i\boldsymbol{\Lambda}t}\bo{S}^\mathrm{T} = \bo{E}^k + i\bo{F}^k \text{.}
\end{align}
We define, 
\begin{subequations} 
        \label{gammaxi} \begin{align}
        ({\boldsymbol{\Gamma}}^k)_{nm} &= \begin{cases} 
        ({\bo{G}}^k)_{nm} \Delta t  &\ n=m\\
        \frac{1}{\lambda_{mn}}\sin (\lambda_{nm}\Delta t )({\bo{G}}^k)_{nm} &\ n \neq m \text{,}
        \end{cases} \\
        ({\boldsymbol{\Xi}}^k)_{nm} &= \begin{cases} 
        0  &\ n=m\\
        \frac{1}{\lambda_{mn}}[1-\cos(\lambda_{nm}\Delta t )]({\bo{G}}^k)_{nm}&\ n \neq m \text{,}
        \end{cases} \end{align}
    \end{subequations}
% \begin{subequations} 
%         \label{gammaxi} \begin{align}
%         ({\boldsymbol{\Gamma}}^k)_{nm} &= \begin{cases} 
%         {\boldsymbol{\Lambda}}^k_{nn} \Delta t  &\ n=m\\
%         -({\bf{S}}^\textrm{T}{\bf{S}}^k)_{nm}\sin (\lambda_{nm}\Delta t ) &\ n \neq m \text{,}
%         \end{cases} \\
%         ({\boldsymbol{\Xi}}^k)_{nm} &= \begin{cases} 
%         0  &\ n=m\\
%         [\cos(\lambda_{nm}\Delta t ) -1]({\bf{S}}^\textrm{T}{\bf{S}}^k)_{nm}&\ n \neq m \text{,}
%         \end{cases} \end{align}
%     \end{subequations}
where $\lambda_{nm} = ({\boldsymbol{\Lambda}})_{mm} - ({\boldsymbol{\Lambda}})_{nn} $, such that,
\begin{align} \label{EandF}
    \bo{E}^k  := \bo{S}^\mathrm{T}\boldsymbol{\Gamma}^k \bo{S} \quad \text{,} \quad
    \bo{F}^k := \bo{S}^\mathrm{T}\boldsymbol{\Xi}^k \bo{S}  \text{,}
\end{align}
where $\bo{E}^k$ is symmetric and $\bo{F}^k$ is skew-symmetric.\cite{Church2018, Cook2023} The nuclear momentum propagation is then, 
\begin{align}
    \label{p-mint-final}
    P_k(t + \Delta t) &= P_k(t) -\left\{V^k_{0} - \frac{\gamma}{2} \textrm{Tr}[\bo{V}^k]\right\} \Delta t - \frac{1}{2} \left\{ \bo{q}^\textrm{T}(t)\bo{E}^k\bo{q}(t) + \bo{p}^\textrm{T}(t)\bo{E}^k\bo{p}(t) - 2\bo{q}^\textrm{T}(t)\bo{F}^k\bo{p}(t)  \right\} \text{.}
\end{align}
The MInt algorithm follows the flow map in Eqn.~\eqref{flow-map-mint}, propagating $\bo{R}$ with Eqn.~\eqref{R-prop-h1} for $H_{1,\textrm{MMST}}$,  $\bo{q}$ and $\bo{p}$ with Eqn.~\eqref{q_p_mint} and $\bo{P}$ with Eqn.~\eqref{p-mint-final} for $H_{2,\textrm{MMST}}$. 
% Further algorithmic details and extension to multiple states is seen in Ref.~[\!\citenum{Church2018}], where we have used a similar notation to Ref.~[\!\citenum{Cook2023}] with the ZPE parameter, $\gamma$. 

\subsection{Symplecticity}
A Hamiltonian system can be described by,\cite{Leimkuhler2005, Church2018, Cook2023}
\begin{align} \label{system}
    \frac{\partial \bo{z}}{\partial t} = \bo{J} \nabla_{\bo{z}} \bo{H} (\bo{z},t) \text{,}
\end{align}
where $\bo{J}$ is the structure matrix and $\bo{z}$ contains the system variables which for the MMST mapping is, $\bo{z}^\mathrm{T} = [\bo{R}^\mathrm{T}, \bo{P}^\mathrm{T}, \bo{q}^\mathrm{T}, \bo{p}^\mathrm{T}]$. In this case where Hamilton's EOM are utilised, 
\begin{align}
    \label{J-MInt}
    \bo{J} = \begin{bmatrix}
        \mathbb{O} & \mathbb{I} \\
        -\mathbb{I} & \mathbb{O}
    \end{bmatrix} \text{,}
\end{align}
where $\mathbb{O}$ and $\mathbb{I}$ are $(F+N)\times(F+N)$ zero and identity matrices respectively. For canonical systems, a Hamiltonian integrator is symplectic if it fulfils the following condition,\cite{Leimkuhler2005, Church2018, Cook2023}
\begin{align}
    \label{symplecticity-eqn}
    \bo{M}^\mathrm{T} \bo{J}^{-1} \bo{M} = \bo{J}^{-1} \text{,}
\end{align}
where $\bo{M}$ is the monodromy matrix defined as, 
\begin{equation}    
  \label{monodromy}
    \bo{M} \equiv \frac{\partial \bo{z}_t}{\partial \bo{z}_0}  \quad\text{where}\quad \bo{M}_{\bo{XY}} = \frac{\partial \bo{X}(t)}{\partial \bo{Y}(0)} \vspace{+5pt}.
\end{equation}
which is a matrix of differentials expressing how the time-evolved phase-space variables depend on the initial phase-space variables.\cite{Irigoyen2006} 
This can be computed for each timestep and multiplied with the previous timestep matrices to obtain the monodromy matrix for the overall trajectory,\cite{Cook2023} 
\begin{gather} 
 \label{monotimestep}
    \bo{M}(2 \Delta t) = \bo{M} (\Delta t \to 2\Delta t ) \bo{M}(0 \to \Delta t).
\end{gather}
For a split Hamiltonian system, $H = H_{1} + H_{2}$, the monodromy matrix is calculated for each timestep according to the algorithmic flow map.\cite{Leimkuhler2005, Cook2023} The MInt algorithm is the only known symplectic algorithm for the MMST Hamiltonian where the criteria in Eqn.~\eqref{symplecticity-eqn} is proven both algebraically and numerically.\cite{Cook2023, Church2018} The MInt algorithm monodromy matrices for one nuclear DoF are shown in Appendix~\ref{Monodromy Matrices} and generalisation to multiple nuclear DoF is in Ref.~[\!\citenum{Church2018}].  

\subsection{Spin-Mapping Representation}
\begin{figure}[t]
        \centering
        \includegraphics[width=0.4\linewidth]{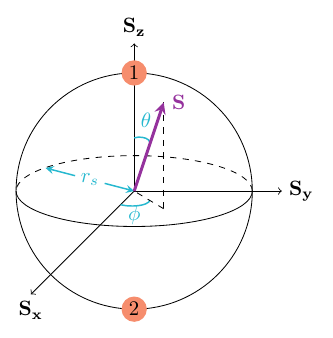}
        \vspace{-10pt}
        \caption{The spin-operator expectation values form a basis where a Bloch sphere of radius, $r_s$, can be defined. The motion of the spin-vector (purple) on the surface is circular around the Hamiltonian vector. When the spin-vector is on either of the poles (orange circle), the population is entirely in one state. }
        \label{fig:spin-mapping}
\end{figure}

Spin-mapping maps a two-level quantum system onto a classical spin-vector on a Bloch sphere by using the equivalence to a spin-1/2 particle in an external magnetic field (which is in a one-to-one correspondence with the Hamiltonian), as seen in Fig.~\ref{fig:spin-mapping}.\cite{Runeson2019} The spin-vector is non-canonical and is propagated using Heisenberg's EOM,\cite{Runeson2019, Bossion2021}      
\begin{align}
    \label{Heisenberg}
    \dot{\bo{S}} = \bo{H} \times \bo{S} \text{,}
\end{align}
where, 
\begin{align} \label{S}
    \bo{S} = \begin{bmatrix}
        \sin\theta\cos\phi\\ \sin\theta\sin\phi \\ \cos\theta
    \end{bmatrix} \text{.}
\end{align}

The quantum-mechanical trace in the correlation function, Eqn.~\eqref{General-CF}, is often calculated by using a Wigner transform, 
\begin{align} \label{wigner}
    A_W(q, p) = \int_{-\infty}^{\infty} dy \, e^{-\frac{i}{\hbar} p y} \left\langle q + \frac{y}{2} \middle| \hat{A} \middle| q - \frac{y}{2} \right\rangle \text{,}
\end{align}
which maps the quantum operators onto a continuous phase-space such that the correlation function can be calculated with classical-like integrals. However, the Wigner transform assumes a Euclidean phase-space which cannot easily be utilised for the spin-vector defined on a spherical phase-space. Instead, the Stratonovich-Weyl transform is used which maps the operators onto a continuous function on the Lie group/manifold, which in the two-state case is the surface of the sphere.\cite{Runeson2019, Runeson2020, Bossion2021, bossionNonadiabaticMappingDynamics2022} There are different transforms; known as Q-, P- and W-functions, which for two states result in radii of $1/2$, $3/2$ and $\sqrt{3}/2$ respectively.\cite{Runeson2019, Bossion2021} Further algebraic details can be found in the literature.\cite{Runeson2019, Runeson2020, bossionNonadiabaticMappingDynamics2022, Bossion2021}

% As the correlation functions tested here are written in the MMST formalism and we sample MMST variables prior to transforming into the spin-variables, we do not require the Stratonovich-Weyl transform explicitly. 

The spin-mapping Hamiltonian is,\cite{Runeson2019} 
\begin{align}
    \label{spin-hamiltonian}
    H_{\textrm{SM}} (\bo{u}) = \frac{1}{2} \bo{P}^\textrm{T}\boldsymbol{\mu}^{-1}\bo{P} + V_{0} + \frac{1}{2} \mathrm{Tr}[\bo{V}] + \frac{1}{2}\bo{H}\cdot \bo{u} \text{,}
\end{align}
where, 
\begin{align}
    \label{H}
    \bo{H} = \begin{bmatrix}
        H_x\\ H_y \\ H_z
    \end{bmatrix} = \begin{bmatrix}
        \Delta^* + \Delta\\
        i(\Delta^* - \Delta)\\ V_1 - V_2
    \end{bmatrix} \text{,}
\end{align}
and we see that $H_x$ is twice the real part of the coupling ($\Delta$) and $H_y$ is twice the imaginary part. We define our non-canonical scaled spin-vector as, 
\begin{align}
    \label{u}
    \bo{u} =  \begin{bmatrix}
        u_x\\ u_y \\ u_z
    \end{bmatrix} = 2r_s \bo{S} = 2r_s\begin{bmatrix}
        \sin\theta\cos\phi\\ \sin\theta\sin\phi \\ \cos\theta
    \end{bmatrix} \text{,}
\end{align}
where $r_s$ is the spin-radius. The spin-mapping Hamiltonian is then,
\begin{align}
    \label{spin-ham-full}
    H_{\textrm{SM}} (\bo{u}) = \frac{1}{2} \bo{P}^\textrm{T}\boldsymbol{\mu}^{-1}\bo{P} + V_{0} + \frac{1}{2} \mathrm{Tr}[\bo{V}] + \frac{1}{2}\left[ 2 \Re(\Delta) u_x + 2 \Im(\Delta) u_y+ (V_1 - V_2)u_z \right] \text{,}
\end{align}
which is equivalent to the ZPE-Reduced MMST Hamiltonian, where $\gamma = 2r_s -1$, due to the following coordinate transform for a two-state system,\cite{Runeson2019} 
\begin{subequations} \label{mmst-spin}
    \begin{align}
        u_x &= q_1q_2 + p_1p_2 \text{,} \\
        u_y &= q_1p_2 - q_2p_1 \text{,}\\
        u_z &= \tfrac{1}{2}(q_1^2 + p_1^2 - q_2^2 -p_2^2) \text{,} 
    \end{align}
\end{subequations}
where $\bo{q}$ and $\bo{p}$ are canonical variables and, 
\begin{align}
    4r_s = q_1^2 + p_1^2 + q_2^2 +p_2^2 \text{.}
\end{align}

Several approaches for integrating the spin-mapping Hamiltonian have been proposed, these include sampling or converting to MMST variables to utilise one of the MMST algorithms\cite{Runeson2019, Runeson2020, Cook2023} and the angle-based algorithm outlined in Appendix~\ref{theta-phi-algo} which we later show is not symplectic.\cite{Bossion2021, bossionNonadiabaticMappingDynamics2022, bossionNonadiabaticRingPolymer2023} Direct angle propagation can become unstable and hence, an extension has been developed to directly propagate the spin-vector elements in regions of instability.\cite{bossionNonadiabaticRingPolymer2023} As far as we are aware, the only suggested symplectic scheme would be using the MInt algorithm.

\section{Methodology} \label{methodology}
We now present the Spin-MInt algorithm and its properties for two electronic DoF. We extend the Spin-MInt algorithm to multiple electronic states in Appendix~\ref{multiple spin mint}.\cite{bossionNonadiabaticMappingDynamics2022, Runeson2020} This algorithm can also be utilised for various path-integral methods by including additional beads.\cite{Hele2015, Hele2013b, Hele2013, Althorpe2013, Richardson2013}

\subsection{Spin-MInt Algorithm} \label{Spin-MInt-algorithm}

We split the spin-Hamiltonian into two sub-Hamiltonians, 
\begin{subequations}
    \begin{align}
        H_{1,\textrm{SM}} &= \frac{1}{2} \bo{P}^\textrm{T}\boldsymbol{\mu}^{-1}\bo{P} \text{,} \\
        H_{2,{\textrm{SM}}} &=  V_0 + \frac{1}{2} \mathrm{Tr}[\bo{V}] + \frac{1}{2}\bo{H}\cdot \bo{u} \text{,}
    \end{align}
\end{subequations}
where despite the spin-vector being non-canonical, the dynamics generated by $H_{2,{\textrm{SM}}}$ using conjugate canonical spin-angle variables (discussed in more detail later and in Ref.~[\!\!\citenum{Bossion2021}]) are identical to Heisenberg's EOM. 
We can define the Spin-MInt algorithm with the following flow map, 
\begin{align}
    \label{flowmap}
    \Psi^{\textrm{Spin-MInt}}_{H_{\textrm{SM}}, \Delta t } := \Phi_{H_{1,{\textrm{SM}}}, \frac{\Delta t}{2}} \circ \Phi_{H_{2,{\textrm{SM}}}, \Delta t} \circ \Phi_{H_{1,{\textrm{SM}}}, \frac{\Delta t}{2}} \text{,}
\end{align}
where only the integration of $H_{2,{\textrm{SM}}}$, $\Phi_{H_{2,{\textrm{SM}}}, \Delta t}$, differs from the MInt algorithm as $H_{1,\textrm{SM}} = H_{1,\textrm{MMST}}$.\cite{Church2018, Cook2023} For $H_{1,{\textrm{SM}}}$, the evolution of $\bo{R}$ is identical to the MInt, Eqn.~\eqref{R-prop-h1}, and all other variables are unchanged. 

For $H_{2,{\textrm{SM}}}$, the EOM are different to the MInt and follow from Eqn.~\eqref{Hamilton's eom} and Eqn.~\eqref{Heisenberg}, 
\begin{subequations}
    \begin{align} \label{p-int-alt}
        \dot{P}_k &= - V^k_0 - \frac{1}{2}\mathrm{Tr}[\bo{V}^k] -\frac{1}{2}\bo{H}^k\cdot \bo{u}(t) \text{,} \\
        \dot{u}_x &= u_z H_y -u_y H_z \text{,} \\
        \dot{u}_y &= -u_z H_x + u_xH_z \text{,}\\
        \dot{u}_z &= u_y H_x - u_x H_y \text{.}
    \end{align}
\end{subequations} 

For the electronic propagation, we can recast the cross product in Eqn.~\eqref{Heisenberg} as a matrix multiplication, 
\begin{subequations} \label{W-evolution}
    \begin{align}
        \dot{\bo{u}} &= \begin{bmatrix}
            0 & -H_z & H_y \\
            H_z & 0 & -H_x \\
            -H_y & H_x & 0 
        \end{bmatrix} \begin{bmatrix}
            u_x \\
            u_y\\
            u_z
        \end{bmatrix} \\
        &= -i \begin{bmatrix}
            0 & -i(V_1 - V_2) & \Delta-\Delta^* \\
        i(V_1-V_2) & 0 & -i(\Delta^* + \Delta) \\
            -(\Delta-\Delta^*) & i(\Delta^* + \Delta) & 0 
        \end{bmatrix} \begin{bmatrix}
            u_x \\
            u_y\\
            u_z
        \end{bmatrix} \\
        &= -i\bo{W}\bo{u} \text{,}
    \end{align}
\end{subequations}
where $\bo{W}$ is Hermitian, skew-symmetric, and all elements are imaginary or zero. We could define this with an entirely real skew-symmetric matrix, but diagonalisation (which we later require) is easiest with a Hermitian matrix. Hence, we define $\bo{W}$ to be $i$ times the skew-symmetric matrix such that the diagonalisation results in real eigenvalues.  

We then integrate to obtain, 
\begin{align}
    \label{u-prop}
    \bo{u}(t+\Delta t) = e^{-i\bo{W}\Delta t}\bo{u}(t) \text{.}
\end{align} 
A time-reversible algorithm recently developed for MASH also has an exponential electronic propagation with a $3\times3$ matrix, but differs in the element construction.\cite{geutherTimeReversibleImplementationMASH2025} We note that as the MASH approach does not have a well-defined Hamiltonian, the Spin-MInt algorithm will not be applicable and symplectic propagation is unlikely to be obtained.\cite{MASH, geutherTimeReversibleImplementationMASH2025}

The momentum integral in Eqn.~\eqref{p-int-alt} integrates as, 
\begin{align}
    \label{p-int-before-dec}
    P_k(\Delta t) 
    &= P_k(0)-\Delta t\left\{V^k_0 + \frac{1}{2}\mathrm{Tr}[\bo{V}^k]\right\} - \frac{1}{2}\int_0^{\Delta t} \textrm{d}t \ \bo{H}^k \cdot e^{-i\bo{W}t}\bo{u}(0) \text{,}
\end{align}
such that we wish to integrate the last term in the `MInt' style. We can decompose $\bo{W}$ into eigenvalues and eigenvectors,  
\begin{align}
    \label{W-eigval-vec}
    \bo{W} = \bo{S}_W \boldsymbol{\Lambda}_W \bo{S}^\dagger_W \text{,}
\end{align}
where as \bo{W} is Hermitian, the conjugate transpose, represented by $\dagger$, is the eigenvector inverse. We define the last term of Eqn.~\eqref{p-int-before-dec} as the integral, $I_u$, and insert $\bo{S}_W\bo{S}^\dagger_W$ identities to obtain, 
\begin{align}
    I_u = \int_0^{\Delta t} \textrm{d}t \ \bo{H}^k \bo{S}_We^{-i\boldsymbol{\Lambda}_Wt}\bo{S}^\dagger_W\bo{u}(0) \text{.}
\end{align}
Element-wise integration results in, 
\begin{align}
    I_u =  \bo{H}^k \bo{S}_W \boldsymbol{\Upsilon}\bo{S}^\dagger_W\bo{u}(0) \text{,}
\end{align}
where, 
\begin{align}\label{upsilon}
    \boldsymbol{\Upsilon}_{nm} = \begin{cases}
         \frac{i}{\lambda_n}(e^{-i\lambda_n \Delta t}-1) &\quad n=m \quad \textrm{and}\quad \lambda_n \neq 0  \\
         \Delta t &\quad n=m \quad \textrm{and}\quad \lambda_n = 0 \\
         0 &\quad n \neq m \text{,}
    \end{cases}
\end{align}
and $\lambda_n$ is the eigenvalue. 
As $\bo{W}$ is a $3\times 3$ Hermitian matrix, one of the eigenvalues will be zero, and the other two are a $\pm \lambda$ pair. This integration is far less complicated than in the MInt algorithm as only one $\boldsymbol{\Upsilon}$ is calculated per timestep (regardless of the number of nuclear DoF) compared to the construction of $F$ symmetric and skew-symmetric $\bo{E}^k$ and $\bo{F}^k$ matrices in Eqns.~\eqref{p_int}--\eqref{EandF}. The integration is similar when generalised to $N$ electronic states, seen in Appendix~\ref{multiple spin mint}, where $\bo{W}$ becomes $N^2-1$ dimensional. Hence, for small $N$ or when $F$ is large (which is common for realistic systems where there are many more nuclear DoF compared to electronic states),\cite{Runeson2020,runeson_exciton_2024,muh_refined_2012, sallerImprovedPopulationOperators2020, Shalashilin2011, Mannouch2022} it is likely that the Spin-MInt will be faster than the MInt algorithm. Numerically, we see that the Spin-MInt is faster for two and three-states, particularly for large $F$, as discussed in Section~\ref{comp-timings}. 

Overall, the integration of $\bo{P}$ is,  
\begin{align} \label{P-h2-prop-alt}
    P_k(t + \Delta t) &= P_k(t) - \Delta t\left(V^k_{0} +  \frac{1}{2}\mathrm{Tr}[\bo{V}^k]\right) - \frac{1}{2} \bo{H}^k \bo{S}_W \boldsymbol{\Upsilon}\bo{S}^\dagger_W\bo{u}(t) \text{,}
\end{align}
which can be implemented analytically. In Section~\ref{mint-eqiuvalence}, we show that trajectories from the Spin-MInt propagation equations are equivalent to the MInt algorithm.

\subsection{Algorithm Steps} \label{algorithm-steps}
The Spin-MInt algorithm for the flow map in Eqn.~\eqref{flowmap} is, 
\begin{enumerate}
    \item Evolve nuclear position for $\Delta t/2$ using Eqn.~\eqref{R-prop-h1}. 
    \item Calculate $\bo{V}(R, \Delta t/2)$, $\bo{V}_0(R, \Delta t/2)$ and derivatives with the updated nuclear position.  
    \item Calculate $\bo{W}(R,\Delta t/2)$ from $\bo{V}(R, \Delta t/2)$ and find $\bo{S}_W$ and $\boldsymbol{\Lambda}_W$ that diagonalise $\bo{W}$.
    \item Evolve $\bo{u}$ for $\Delta t$ using Eqn.~\eqref{u-prop}.
    \item Use $\boldsymbol{\Lambda}_W$ to calculate $\boldsymbol{\Upsilon}$. Find the derivative of $\bo{H}$. 
    \item Evolve nuclear momentum for $\Delta t$ with Eqn.~\eqref{P-h2-prop-alt}.
    \item Repeat step 1 for the second half-timestep of $H_{1, \textrm{SM}}$. 
\end{enumerate}
such that repeating for $n$ timesteps results in a simulation of length $n\Delta t$ time units.

\subsection{Spin-MInt Properties} \label{properties}

We wish to prove that the Spin-MInt algorithm is a symplectic, symmetric, second-order, time-reversible, angle invariant and structure preserving method. Several of these properties can be obtained through showing that the MInt and Spin-MInt propagations are equivalent.

\subsubsection{Equivalence to the MInt Algorithm}
\label{mint-eqiuvalence}

We show the Spin-MInt is equivalent to the MInt by comparing the $H_2$ integrations. $u_x$ can be expressed as,\cite{Runeson2020}
\begin{align}
    u_x &= (\mathbf{q}-i\mathbf{p})^{\text{T}}\hat{S}_1(\mathbf{q}+i\mathbf{p}) \text{,}
\end{align}   
where $\hat{S}_1 = \frac{1}{2}\bigl[\begin{smallmatrix} 0 &1\\1&0\end{smallmatrix}\bigr]$ is half of the first Pauli spin matrix and from Eqn.~\eqref{q_p_prop},\cite{cookElectronicPathIntegral2025, Hele2016, Chowdhury2021}
\begin{equation}
    \mathbf{\dot{q}}+i\mathbf{\dot{p}} = -i\mathbf{V}(\mathbf{q}+i\mathbf{p}) \text{,}
\end{equation}
such that, 
\begin{align}
    \label{u_xEvol}
    \dot{u}_x = i(\mathbf{q}-i\mathbf{p})^\text{T}[\mathbf{V},\hat{S}_1](\mathbf{q}+i\mathbf{p}) \text{.}
\end{align} 
Decomposing $\mathbf{V}$ into traceless and identity-like components where, 
\begin{align}
\mathbf{V} = \begin{bmatrix} \frac{V_1-V_2}{2} &\Delta^* \\ \Delta & \frac{V_2-V_1}{2}
\end{bmatrix}  + \frac{V_1+V_2}{2}\mathbb{I} \text{.}
\end{align}
The commutator is then,
\begin{align}
    \label{PauliAntiCommutator}
    [\mathbf{V},\hat{S}_1] = i\left(V_1-V_2\right) \hat{S}_2 + (\Delta^*-\Delta) \hat{S}_3 \text{,}
\end{align}
where $\hat{S}_2$ and $\hat{S}_3$ are half of the second and third Pauli matrices respectively. Substituting into Eqn.~\eqref{u_xEvol} and repeating for $u_y$ and $u_z$ results in Eqn.~\eqref{W-evolution} which we integrate exactly. Therefore, the MInt and Spin-MInt electronic evolutions are equivalent.

We now do the same for the nuclear momentum propagation, comparing the integrated forms of $P$ as alternative integration methods have been shown to produce different results in previous work.\cite{Cook2023} The Spin-MInt propagation of $\bo{P}$ in Eqn.~\eqref{P-h2-prop-alt} is equivalent to,
\begin{align} \label{P-h2-prop}
    (P_k)_{\textrm{SM}}(t + \Delta t) &= P_k(t) - \Delta t\left(V^k_{0} +  \frac{1}{2}\mathrm{Tr}\left[\bo{V}^k\right]\right) - \frac{1}{2} \mathrm{Tr}\left[\left(\bo{E}^k + i\bo{F}^k \right)\bo{C}(\bo{u},t)\right] \text{,}
\end{align}
as outlined in Section IC of the Supplementary Material where, 
\begin{align}
    \label{C-u}
    \bo{C}(\bo{u}) = \begin{bmatrix}
        u_z & u_x -iu_y \\
        u_x + iu_y & -u_z
    \end{bmatrix} \text{,}
\end{align}
which is related to the $\bo{C}(\bo{q}, \bo{p}) =  (\bo{q} + i \bo{p}) \bigotimes (\bo{q} - i\bo{p})^\mathrm{T}$ matrix,\cite{cookElectronicPathIntegral2025, Hele2016}
\begin{align}
    \bo{C}(\bo{q}, \bo{p})
    &= \bo{C}(\bo{u}) + 2r_s\mathbb{I} \text{,}
\end{align}
such that we obtain the MInt $\bo{P}$ propagation, Eqn.~\eqref{p-mint-final},
\begin{subequations} \label{equivPProp}
    \begin{align}
        (P_k)_{\text{SM}}(t+\Delta t) 
        &=P_k(t) - \Delta t \left( V^k_0 +\frac{1}{2} \textrm{Tr}\left[\mathbf{V}^k\right] \right)  - \frac{1}{2}\textrm{Tr}\left[\left(\mathbf{E}^k+i\mathbf{F}^k \right) (\mathbf{C}(\mathbf{q},\mathbf{p},t) - 2r_s\mathbb{I})\right] \\
        &= P(t) - \Delta t\left( V^k_0 - \frac{\gamma}{2}\textrm{Tr}\left[\mathbf{V}^k\right] \right) - \frac{1}{2}\textrm{Tr}\left[\left(\mathbf{E}^k+i\mathbf{F}^k\right)\mathbf{C}(\mathbf{q},\mathbf{p},t)\right] \text{,}
    \end{align}
\end{subequations}
due to the symmetry of $\mathbf{E}^k/\mathbf{F}^k$ and $\textrm{Tr}[\mathbf{E}^k] = \textrm{Tr}(\mathbf{V}^k) \Delta t $.\cite{Church2018} Hence, both algorithms are equivalent. We have chosen to propagate Eqn.~\eqref{P-h2-prop-alt} in the Spin-MInt algorithm which removes the need to calculate $\boldsymbol{\Gamma}^k/\boldsymbol{\Xi}^k$ to obtain $\mathbf{E}^k/\mathbf{F}^k$, reducing the computational cost.

As the same flowmap is used for the MInt and Spin-MInt algorithms, it can then easily be seen that the Spin-MInt will be symmetric, second-order, symplectic and time-reversible due to the same reasoning as for the MInt algorithm.\cite{Church2018} However, as we do not calculate several matrices needed to prove symplecticity numerically within the Spin-MInt propagation, we have chosen to show symplecticity without reliance on the equivalence to the MInt.

\subsubsection{Symplecticity}

As the spin-vector is non-canonical, the structure matrix for evaluation of the symplecticity criterion in Eqn.~\eqref{symplecticity-eqn} is ill-defined due to the cross-dependence of the elements. Previous work has suggested defining a non-canonically symplectic integrator, or more commonly referred to as a Poisson integrator, as one that is structure preserving where the spin-magnitude is conserved throughout simulation.\cite{hairerGeometricNumericalIntegration2006,lubichSymplecticIntegrationPostNewtonian2010} In section~\ref{sp-ai}, we show that this is not sufficient to prove symplecticity when there is non-trivial coupling between canonical and spin-systems. Hence, we would like to prove that this algorithm is rigorously symplectic which can numerically be tested using the Spin-MInt trajectories. 

To retain the $2N-2$ DoF associated with the spin-mapping, we utilise a transformation into canonical conjugate variables to re-obtain Hamilton's EOM.\cite{Bossion2021} These conjugate variables are linked to the spin-vector through Eqn.~\eqref{u} and are, 
\begin{subequations} \label{conj-var}
    \begin{align}
        \dot{\phi} &= \frac{\partial H_{2,{\textrm{SM}}}}{\partial r_s\cos\theta} \text{,} \\
        \dot{r_s\cos\theta} &= -\frac{\partial H_{2,{\textrm{SM}}}}{\partial \phi} \text{,}
    \end{align}
\end{subequations}
which results in the same structure matrix as the MMST representation, Eqn.~\eqref{J-MInt}, where the identity and zero matrices are $(F+1)\times(F+1)$. The system is then described by $\bo{z}^\mathrm{T} = [\bo{R}^{\textrm{T}}, \phi, \bo{P}^{\textrm{T}}, r_s \cos\theta]$ such that the monodromy matrix is, 
\begin{align}
    \bo{M}_{\textrm{SM}} = \begin{bmatrix}
        \textrm{M}_{\bo{R}\bo{R}} & \textrm{M}_{\bo{R}\phi} &\textrm{M}_{\bo{R}\bo{P}} &\textrm{M}_{\bo{R}r_s\cos\theta} \\
        \textrm{M}_{\phi \bo{R}} &\textrm{M}_{\phi\phi} &\textrm{M}_{\phi \bo{P}} &\textrm{M}_{\phi r_s\cos\theta} \\
        \textrm{M}_{\bo{P}\bo{R}} &\textrm{M}_{\bo{P}\phi} &\textrm{M}_{\bo{P}\bo{P}} &\textrm{M}_{\bo{P} r_s\cos\theta} \\
        \textrm{M}_{r_s\cos\theta \bo{R}} &\textrm{M}_{r_s\cos\theta \phi} &\textrm{M}_{r_s\cos\theta \bo{P}} &\textrm{M}_{r_s\cos\theta r_s\cos\theta } 
    \end{bmatrix} \text{.}
\end{align}

For simplicity, we define the monodromy matrix elements for the Spin-MInt algorithm in Appendix~\ref{symp-spin-mint} for one nuclear DoF. Further nuclear DoF will add indices as seen in Ref.~[\!\citenum{Church2018}]. In Section IE of the Supplementary Material, we show satisfaction of Liouville's theorem ($|\bo{M}| = 1$) which results in phase-space preservation. In Section ID of the Supplementary Material, we define an explicit version of the monodromy matrix for two states where $H_y =0$, which is equivalent to the form in Appendix~\ref{symp-spin-mint} but challenging to generalise for more than two electronic states.
Whilst we can infer symplecticity from the exact sub-Hamiltonian evolution, we algebraically prove symplecticity in Section IF of the Supplementary Material for two states with $H_y = 0$ using the explicit form of the monodromy matrix. Time-reversibility follows from this coordinate transform and is outlined in the Supplementary Material Section G. 

The Spin-MInt is therefore, to the best of our knowledge, the first known symplectic algorithm for a spin-mapping Hamiltonian where symplecticity is rigorously proven using a canonical coordinate transform into the angle description of the spin-vector.  

\subsubsection{Structure Preservation and Angle Invariance} \label{sp-ai}
We can show that the Spin-MInt algorithm preserves the geometric structure of the Poisson bracket through the Casimirs' of the bracket,\cite{luesinkCasimirPreservingStochastic2024, lubichSymplecticIntegrationPostNewtonian2010} which results in conservation of the magnitude of $\bo{u}$ and further details are in Ref.~[\!\!\citenum{Runeson2020}]. The spin-vector is evolved under Eqn.~\eqref{u-prop} such that, 
\begin{align}
    \label{transformation}
    \bo{T} = e^{-i\bo{W} \Delta t} \text{,}
\end{align}
is the transformation matrix which is unitary as $\bo{W}$ is Hermitian. The norm of the spin-vector is preserved as, 
\begin{align}
    ||\bo{T}\bo{u}||^2 = (\bo{T}\bo{u})^\dagger(\bo{T}\bo{u}) = ||\bo{u}||^2 \text{,}
\end{align}
such that, $\braket{\bo{u}}{\bo{u}}$ is conserved. This is shown numerically in the Supplementary Material Figure S.10. 
Structure preservation is a weaker criteria than symplecticity, as one can derive a structure preserving but not symplectic spin-mapping algorithm utilising a Split-Liouvillian (SL) formalism. We algebraically show this in Appendix~\ref{Spin-SL-algo} through the derivation of the spin Split-Liouvillian (Spin-SL) algorithm where we obtain a structure preserving electronic evolution but not exact sub-Hamiltonian $H_{2, \textrm{SM}}$ evolution.

Structure preservation also results in an invariance to the overall phase of the spin-vector. A phase-space transformation is described by,
\begin{align} \label{angle-shift}
    \tilde{\bo{u}} = e^{-i\theta}\bo{u} \text{,}
\end{align}
which commutes with $\mathbf{T}$ for any scalar $\theta$. Therefore, the spin-vector evolution is unaffected by any angle shift using Eqn.~\eqref{angle-shift} and back again.

\section{Models} \label{models}

For algorithmic comparison, we firstly utilise the same two-state model (one-dimensional spin-boson model) used in Ref.~[\!\citenum{Cook2023}] to extend our benchmarking to spin-mapping algorithms.\cite{Richardson2013} The diabatic potential matrix is, 
\begin{align} \label{potentialmatrix}
    \tilde{\bo{V}} &= V_0 + \bo{V} \\
    &= \frac{1}{2}m\omega^2R^2 + \left[ {\begin{array}{cc} \alpha +\kappa R & \Delta \\
    \Delta & -\alpha - \kappa R \end{array}} \right] \text{,}
\end{align}
where there are three models (1, 2, 3) tabulated in the Supplementary Material (Table S.2) for coupling, $\Delta$, and asymmetry, $\alpha$. The potential is split into a traceless state-dependent matrix, $\bo{V}$, and a state-dependent potential, $V_0$, to allow literature comparison but neither algorithm requires a traceless $\bo{V}$ matrix.\cite{Cook2023} For this model, the one-bead NRPMD distribution is sampled as in Appendix D of Ref.~[\!\citenum{Cook2023}] (and in the Supplementary Material Section IIA) to directly extend the previous benchmarking. Model 1 is in the adiabatic limit where there is strong electronic coupling. Model 2 is in the inverted Marcus regime where there is a large energy bias between surfaces. Model 3 is an intermediate regime which is the most challenging to simulate due to similar timescales of electronic and nuclear dynamics.\cite{Cook2023} The results for model 1 are shown here, models 2 and 3 can be seen in the Supplementary Material in Section IV. 

We extend to more nuclear modes by considering a spin-boson model coupled to a bath of harmonic modes.\cite{Runeson2019, cotton_new_2016} This system-bath model has the following potential matrix for $F$ bath modes, 
\begin{align} \label{potentialmatrix-spinboson}
    \tilde{\bo{V}} 
    % &= \bo{V}_0 + \bo{V} \\
    &= \sum_j^F \frac{1}{2}m_j\omega_j^2R_j^2 + \left[ {\begin{array}{cc} \epsilon + \sum_j^F c_j R_j & \Delta \\
    \Delta & -\epsilon - \sum_j^F c_j R_j \end{array}} \right] \text{,}
\end{align}
where there are four models (I, II, III, IV) with parameters tabulated in the Supplementary Material (Table S.3). $\omega_j$ and $c_j$ are calculated from the discretisation of the spectral density as outlined in Refs.~[\!\citenum{Runeson2019, craig_chemical_2005}] with $F=100$ bath modes. The electronic variables are sampled from a hypersphere of radius $2$ and the nuclear variables from a thermal Wigner distribution, as outlined in the Supplementary Material Section IIB. Models I and II are a symmetric potential at low and high temperatures respectively and models III and IV are an asymmetric potential at low and high temperatures.\cite{Runeson2019, craig_chemical_2005}

Finally, we use a three-state system with one nuclear DoF, a Morse potential model,\cite{bossionNonadiabaticMappingDynamics2022, coronadoUltrafastNonadiabaticDynamics2001} with the potential matrix,
\begin{subequations} \label{morse-pot}
\begin{align}
    V_{ii} &= D_{ii}\left( 1- e^{-\alpha_{ii}(R - R_{ii})} \right)^2 + c_{ii} \text{,} \\
    V_{ij}& = A_{ij}e^{-\alpha_{ij}(R - R_{ij})^2} \text{,} 
\end{align}
\end{subequations}
where we have three models (A, B, C) with tabulated parameters in the Supplementary Material (Table S.4). For this model, the electronic sampling is focused on the first state and we sample a nuclear Wigner distribution, as detailed in the Supplementary Material Section IIC.\cite{bossionNonadiabaticMappingDynamics2022} This is an anharmonic model and by computing population dynamics we model photo-dissociation.\cite{bossionNonadiabaticMappingDynamics2022} 

The sampled distributions are tabulated in the Supplementary Material (Table S.5), where we additionally include the relevant figure numbers for clarity. The one-dimensional spin-boson model samples the NRPMD distribution and calculates MMST observables for benchmarking purposes. In contrast, we utilise spin-mapping sampling and observables for the system-bath model and three-state Morse potential. 

\section{Results}\label{results}
Here, we present single trajectory and ensemble trajectory properties for the one-dimensional spin-boson model. We show the time-dependent population difference, $\sigma_z$, for the system-bath model with 100 nuclear modes, and the state populations for the three-state Morse potential model. We also discuss algorithmic computational timings for the system-bath model and the three-state Morse potential. 

\subsection{Single Trajectory Comparison}

\begin{figure}[h]
    \centering
    \includegraphics[width=0.45\linewidth]{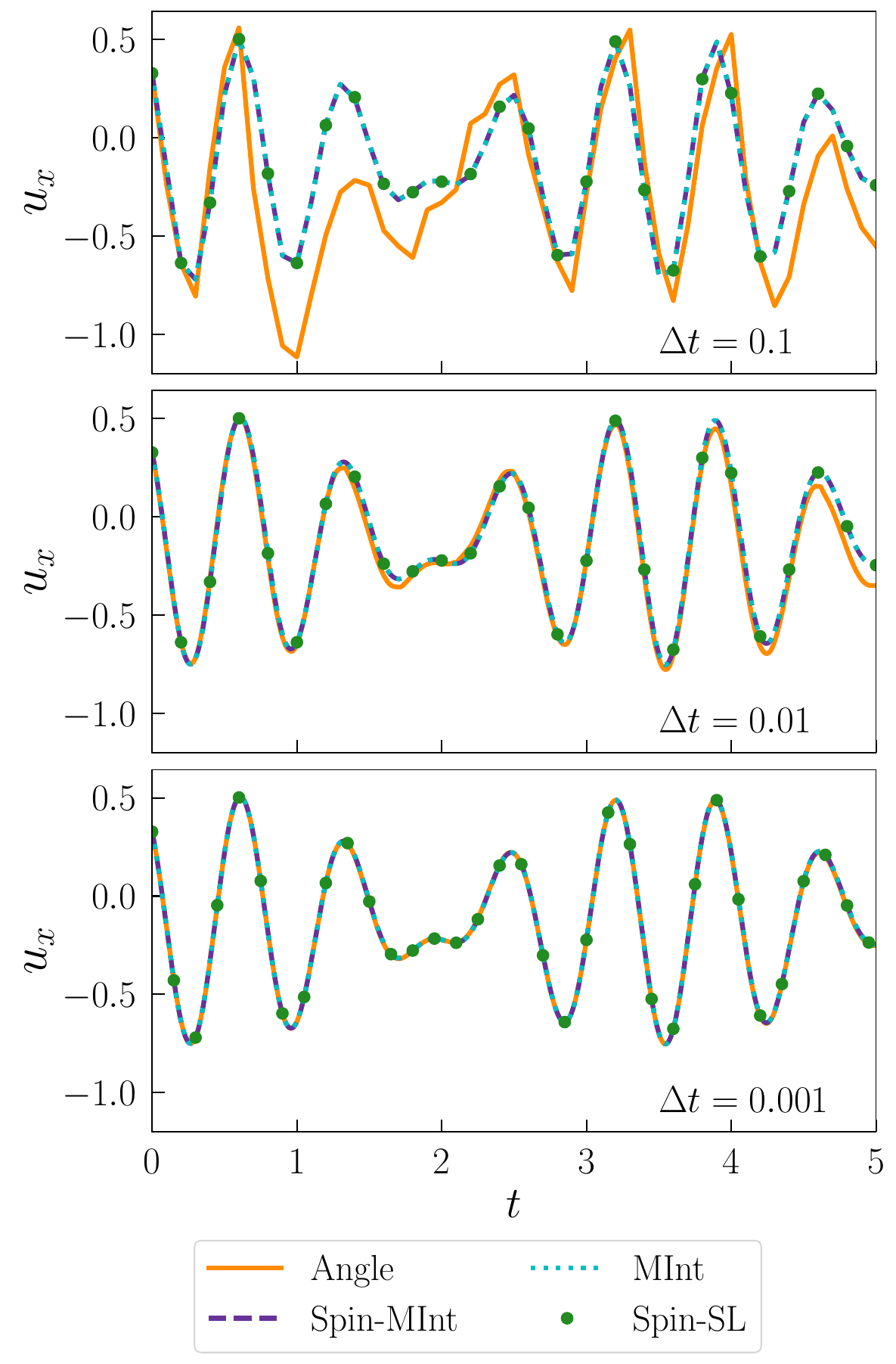}
    \vspace{-10pt}
    \caption{Dynamics of $u_x$ for a single trajectory using model 1 and the MInt (cyan dotted), Spin-MInt (purple dashed), angle-based (orange) and Spin-SL (green circles) algorithms at three different timesteps: 0.1 (top), 0.01 (middle) and 0.001 (bottom). The Spin-MInt, MInt and Spin-SL results are identical at all timesteps whilst the angle-based algorithm deviates for the 0.1 and 0.01 timesteps indicating lower accuracy.}
    \label{fig:singletraj30}
\end{figure}

For a single trajectory, we compare the accuracy of the Spin-MInt algorithm against the MInt algorithm, Spin-SL algorithm and an angle-based algorithm from Ref.~[\!\citenum{Bossion2021}].\cite{Church2018, Bossion2021} The angle-based algorithm directly propagates $\theta$ and $\phi$ and is outlined in Appendix~\ref{theta-phi-algo}.\cite{Bossion2021, bossionNonadiabaticMappingDynamics2022} In Figure~\ref{fig:singletraj30}, we show the propagation of $u_x$ using the four algorithms. We note that for symmetric propagation the Spin-SL uses the alternative sub-Hamiltonian ordering, Eqn.~\eqref{SLflow2}. We find that a small timestep is required to obtain accurate propagation using the angle-based algorithm. However, the Spin-MInt, MInt and Spin-SL algorithms are more tolerant of coarser timesteps, capturing the correct dynamics at all timesteps in Figure~\ref{fig:singletraj30}. While the Spin-MInt and MInt algorithms are identical in propagation algebraically, the Spin-SL appears to also produce the same trajectories as seen in the SI for nuclear and electronic variables with all three models (Figure S.1-S.6). As the Spin-MInt, MInt and Spin-SL algorithms have equal electronic evolution, the Spin-SL will only differ on nuclear momentum propagation which may affect the trajectories at longer simulation times.

\begin{figure}[b]
    \centering
    \includegraphics[width=0.45\linewidth]{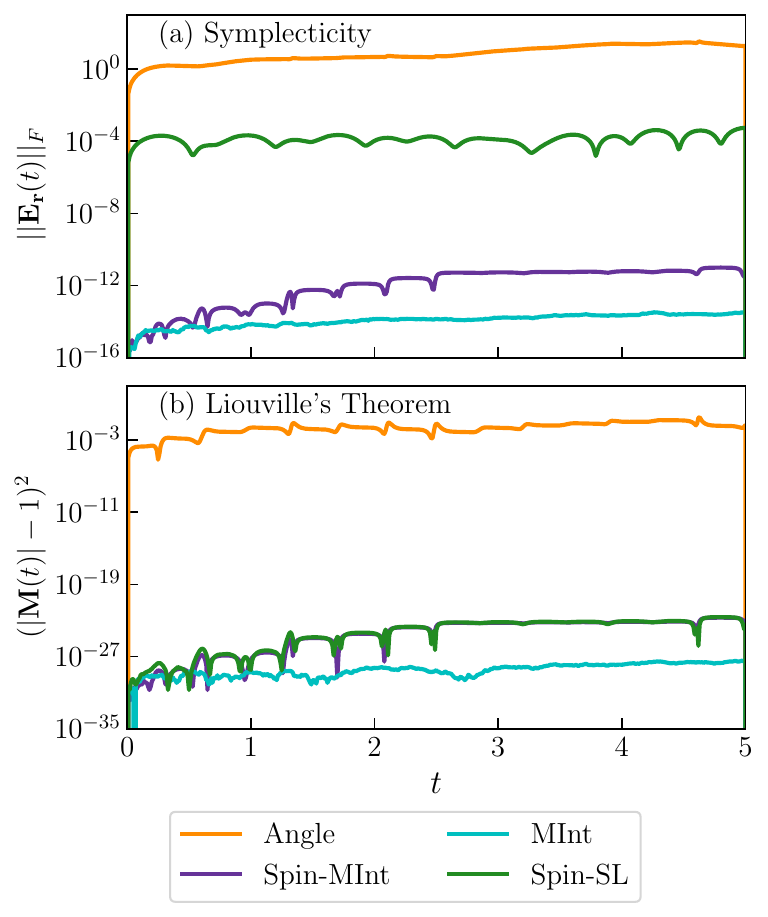}
    \vspace{-10pt}
    \caption{(a) The error matrix Frobenius norm and (b) the Liouville's theorem criterion as a function of time using model 1 and $\Delta t = 0.01$, for a single trajectory using the Spin-MInt (purple), MInt (cyan), angle-based (orange) and Spin-SL (green) algorithms. As the MInt and Spin-MInt algorithms measure floating point error in both plots, Liouville's theorem and symplecticity are satisfied. The Spin-SL algorithm satisfies Liouville's theorem but is not symplectic. The angle-based algorithm does not satisfy either condition. }
    \label{fig:sympcompare}
\end{figure}

It has been noted in the literature that when $\theta \to 0, n \pi$, angle-based algorithms are likely to become unstable and swapping to propagation of the spin-vector is advised.\cite{Bossion2021, bossionNonadiabaticMappingDynamics2022} However, as this makes computing the monodromy matrix complicated, to compare the symplecticity and satisfaction of Liouville's theorem, we focus on trajectories that do not reach these unstable regions. In Figure~\ref{fig:sympcompare}, we present the symplecticity criterion and satisfaction of Liouville's theorem for the trajectory in Figure~\ref{fig:singletraj30} using model 1 with $\Delta t =0.1$. 

We evaluate the symplecticity criterion by defining an error matrix,\cite{Cook2023}
\begin{align} \label{errormatrix}
    \bo{E_{r}} &= \bo{M}^\textrm{T}\bo{J}^{-1}\bo{M} \ -  \ \bo{J}^{-1} \text{,}
\end{align}
where for a symplectic integrator the elements of $\bo{E_{r}}$, $e_{ij}$, will be zero.\cite{Ivanov2013}
We then utilise the Frobenius Norm to track the size of the error matrix,\cite{Cook2023, Ivanov2013}
\begin{align} 
    \label{frobnorm}
    ||{\bo{E_{r}}}||_{F} &= \sqrt{\sum_{i=1}^n\sum_{j=1}^n |e_{ij}|^2} \text{.}
\end{align}
In Figure~\ref{fig:sympcompare} (a), we show a logarithmic plot of $||{\bo{E_{r}}}||_{F}$ against time for the single trajectory in Figure~\ref{fig:singletraj30}. The Spin-MInt (purple) and MInt (cyan) algorithms remain below $10^{-12}$ for the whole trajectory and therefore are symplectic. However, the angle-based algorithm (orange) and Spin-SL (green) rapidly increase, indicating lack of symplecticity.

We can define two mathematically identical ways of implementing the Spin-MInt algorithm for the two-state potential tested in Eqn.~\eqref{potentialmatrix}. To calculate $e^{-i\bo{W}t}$, one can decompose into eigenvalues and eigenvectors (resulting in the monodromy matrix in Appendix~\ref{symp-spin-mint}) or through an explicit form outlined in the Supplementary Material Section ID. The explicit form of results in lower floating point error, seen in the Supplementary Material Figure S.8, but is challenging to obtain for multiple nuclear and electronic DoF such that the monodromy in Appendix~\ref{symp-spin-mint} is used here. 

To evaluate satisfaction of Liouville's theorem, where $|\bo{M}|=1$,\cite{Cook2023, Hardy1995, Tuckerman2010} we plot $(|{\bf{M}}|-1)^2$ in Figure~\ref{fig:sympcompare} (b). Satisfying Liouville's theorem does not always indicate an accurate algorithm as it is equivalent to having a divergenceless Liouvillian.\cite{Ezra2004, Cook2023} We see that the Spin-MInt, MInt and Spin-SL algorithms satisfy Liouville's theorem. The angle-based algorithm does not satisfy Liouville's theorem which is uncommon for a mapping algorithm.\cite{Cook2023} From the algebraic monodromy in Appendix~\ref{theta-phi-algo}, it is clear that the determinant will initially deviate from unity on $\mathcal{O}(\Delta t)$ as in Eqn.~\eqref{theta-phi-det} such that it is zero-order with respect to timestep when propagating for a given length of time.\cite{Cook2023} In the Supplementary Material Figure S.7, we show that the angle-based algorithm is zero-order with respect to timestep whilst the Spin-SL algorithm is second-order.  

For a single trajectory, the Spin-MInt and MInt algorithms have identical propagation and are symplectic. The Spin-SL algorithm has accurate trajectories and satisfies Liouville's theorem, in contrast to the angle-based algorithm which requires a small timestep to accurately capture the dynamics. As neither the Spin-SL or angle based algorithms are symplectic, we do not consider them further in our search for a symplectic spin-mapping algorithm. However, extending the Spin-SL algorithm to multiple electronic states and testing the ensemble properties is an interesting avenue for future work.  

\subsection{Ensemble Properties}

Ensemble properties were tested by averaging over many trajectories using the Spin-MInt and MInt algorithms with the one-dimensional spin-boson model. Both algorithms are symplectic and satisfy Liouville's theorem, seen in the Supplementary Material Figure S.9. The Spin-MInt is also numerically structure preserving, seen by testing the spin-magnitude conservation in Figure S.10. 

To investigate energy conservation, the criterion is,\cite{Cook2023}
\begin{equation} \label{energyequation}
    \langle\left[\epsilon(t) - \epsilon(0)\right]^2\rangle_{\rho} = \frac{\sum_{i=1}^J \left[\epsilon(t) - \epsilon(0)\right]^2 Q_{i}(0)}{\sum_{i=1}^J Q_{i}(0)} \text{,}
\end{equation} 
where the energy, $\epsilon$, is calculated by evaluating the Hamiltonians, Eqn.~\eqref{full-MMST} and Eqn.~\eqref{spin-ham-full} for the MInt and Spin-MInt algorithms at each timestep.\cite{Cook2023} The weighting is $Q = {\bf{p}}^\textrm{T}e^{-\beta{{\bo{V}}}/2}{\bf{q}}\times{\bf{q}}^\textrm{T}e^{-\beta{{\bo{V}}}/2} {\bf{p}}$.

\begin{figure}[h]
        \centering
        \includegraphics[width=0.5\linewidth]{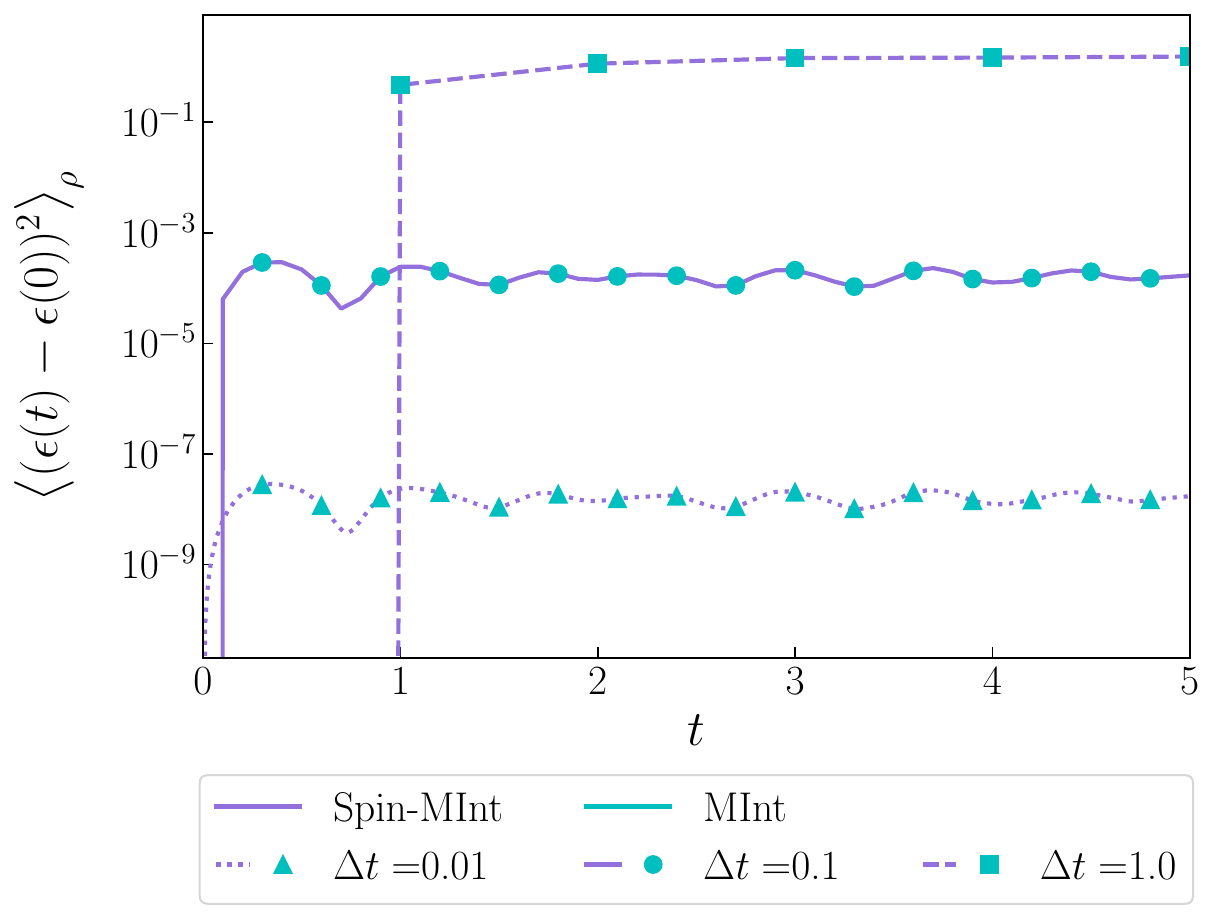}
        \vspace{-10pt}
        \caption{Energy conservation of model 1 averaged over trajectories using $\Delta t = 0.1$ (solid/circle), $\Delta t = 0.01$ (dotted/triangle) and $\Delta t = 1.0$ (dashed/square) with the Spin-MInt (purple) and MInt (cyan) algorithms. The energy is the same for both algorithms (due to the identical trajectories) and is second-order.}
        \label{fig:Econs}
\end{figure}

In Figure \ref{fig:Econs}, we plot the energy conservation for an ensemble of trajectories. Due to the identical trajectories, the trajectory energy is the same such that both algorithms are second-order. Numerically, we see this as changing the timestep by a factor of $10$ results in a change of $10^4$ in the energy criterion.\cite{Cook2023} This agrees with the algebraic determination of order for the MInt in Ref. [\!\!\citenum{Church2018}] and the derivation in Section \ref{properties} for the Spin-MInt.\cite{Church2018} For a symplectic algorithm, this means the energy of the approximate Hamiltonian is exactly conserved which differs from the exact Hamiltonian by the order of the algorithm,\cite{Leimkuhler2005, Church2018, Cook2023} resulting in energy conservation at exponentially long-times with fluctuations on $\mathcal{O}(\Delta t^2)$.\cite{Cook2023}

We present the nuclear position and electronic population NRPMD autocorrelation functions in the Supplementary Material Figure S.12 with $\Delta t = 0.1$. Both algorithms reproduce the same correlation functions as in the literature, indicating accurate propagation.\cite{Cook2023} We also sample spin-mapping variables and calculate the one bead SM-NRPMD correlation function (outlined in Ref.~[\!\citenum{Bossion2021}]). We highlight the following advantages of the spin-mapping approach over MMST noting that both algorithms can be used for either representation. We find convergence with at least five times fewer trajectories compared to NRPMD as sampling a sphere is much more efficient than Cartesian space. The SM-NRPMD electronic correlation functions agree with NRPMD for the adiabatic limit of model 1 but captures the zero-time behaviour of models 2 and 3 more accurately compared to the exact result in the literature.\cite{Bossion2021} 

\subsection{System-Bath Results}
\begin{figure}[t]
        \centering
        \includegraphics[width=0.9\linewidth]{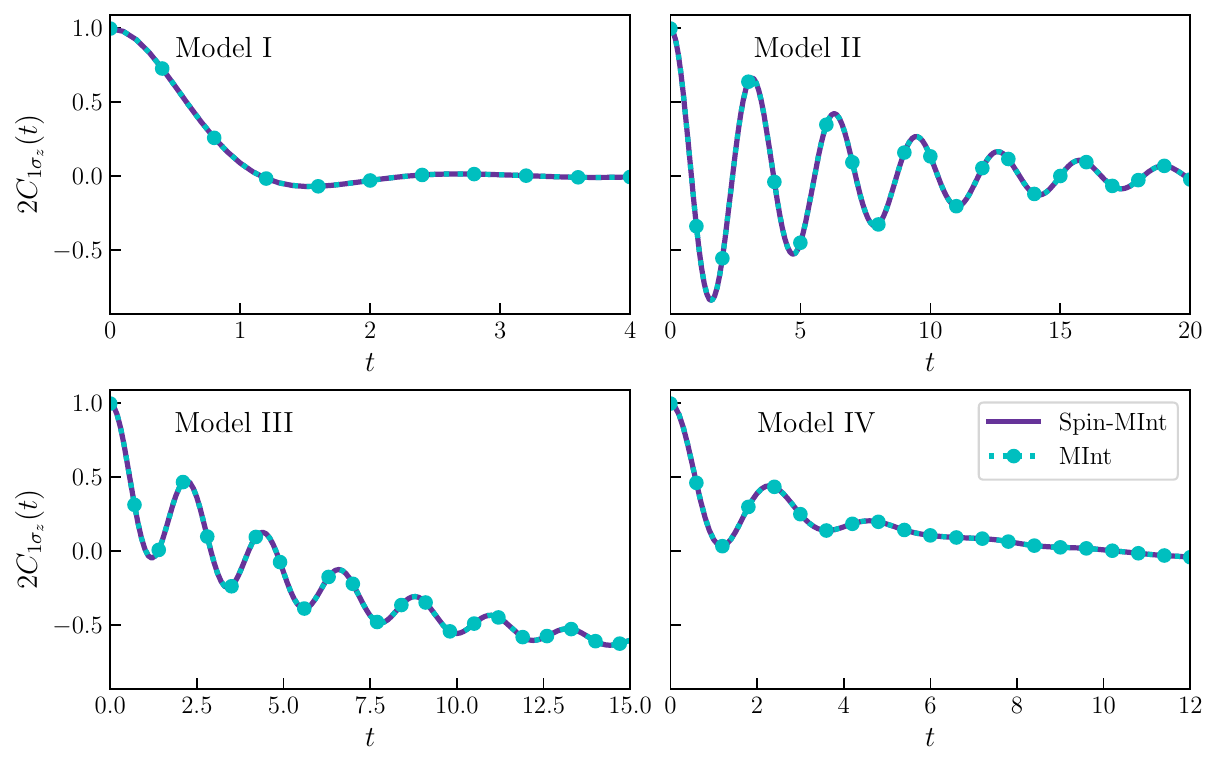}
        \vspace{-20pt}
        \caption{Correlation function for the time-dependant population difference after initial excitation to state 1 using the system-bath models with the Spin-MInt (purple) and MInt (cyan dotted circles) with $F=100$ nuclear modes. The results were calculated using $10^5$ trajectories with a timestep of $0.1$. The Spin-MInt and MInt give the same result in agreement with the literature. }
        \label{fig:spinboson}
\end{figure}
We compute the time-dependent population difference after initial excitation to state 1, $C_{1\sigma_z}$, for the system-bath model with $F =100$ nuclear modes.\cite{Runeson2019} The correlation function is
\begin{align}
    C_{1\sigma_z} = \left\langle  \frac{1}{2}\left[q_1^2 (0) +p_1^2(0) \right] \times \sigma_z(t) \right\rangle_\rho \text{,}
\end{align}
where we average over the sampled distribution, $\rho$. In the spin-mapping representation, 
\begin{align}
    \sigma_z = u_z \times \frac{r_s}{r_{\bar{s}}}
\end{align}
and the MMST representation, 
\begin{align}
    \sigma_z = \frac{1}{2}(q_1^2 + p_1^2 - q_2^2 -p_2^2 ) \times \frac{r_s}{r_{\bar{s}}}
\end{align}
where $r_s = 1/2$ and $r_{\bar{s}}= 3/2$ such that we calculate the $Q$ method in Ref.~[\!\citenum{Runeson2019}]. In Figure~\ref{fig:spinboson}, we see that the MInt and Spin-MInt results agree and we encourage comparison to the exact result in Refs.~[\!\citenum{Runeson2019, craig_chemical_2005}]. Whilst this model highlights that the extension to multiple nuclear states is accurate, the main benefit is investigating the computational scaling with nuclear DoF as discussed later.

\subsection{Three-State Results}
\begin{figure}[t]
        \centering
        \includegraphics[width=\linewidth]{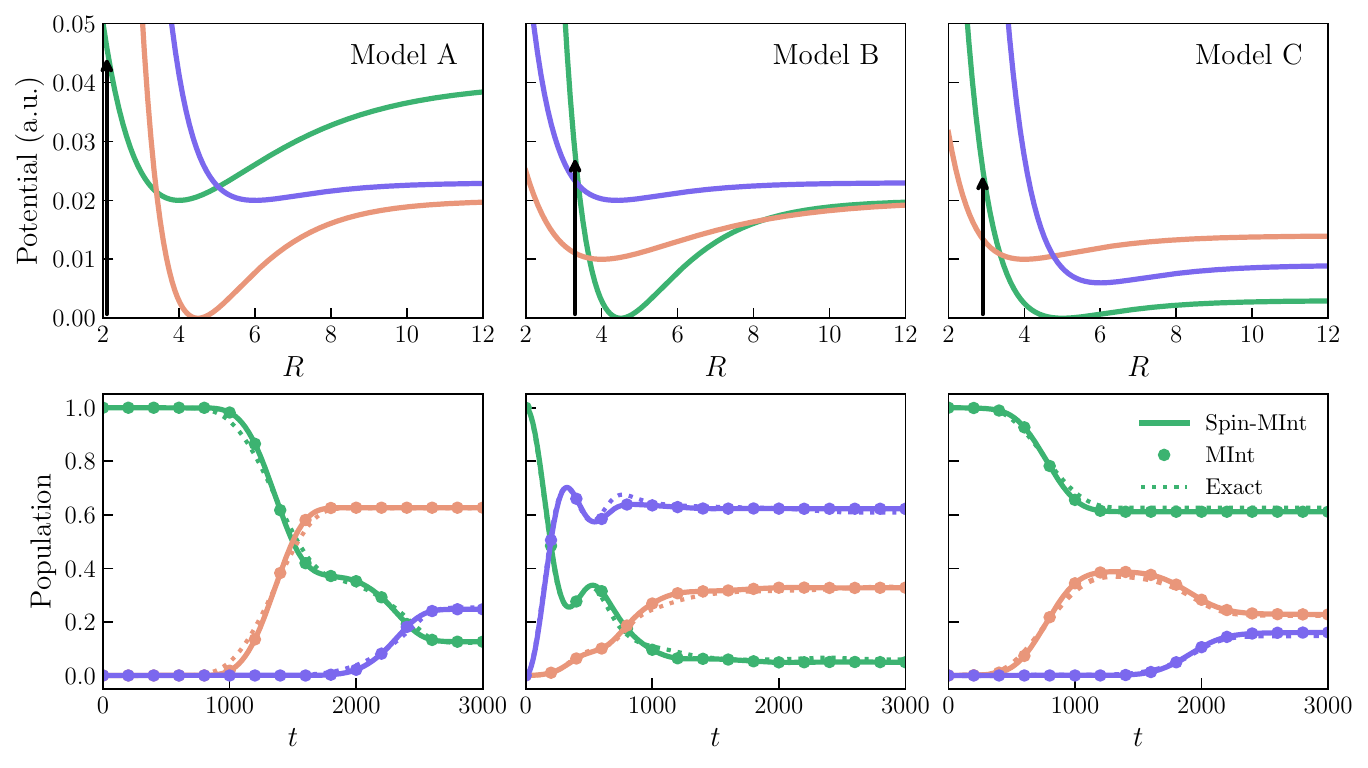}
        \vspace{-40pt}
        \caption{Three-state Morse potential results for models A, B and C with the diabatic potential energy surfaces (top row) and state populations (bottom row). Focused initial conditions are on the first state (green) at the vertical arrows. The MInt (circles) and Spin-MInt (solid line) result in identical population dynamics that agree closely with the exact result (dotted).}
        \label{fig:morse}
\end{figure}
In Appendix~\ref{multiple spin mint}, we generalise the Spin-MInt algorithm to multiple electronic states. In Figure~\ref{fig:morse}, we plot the population dynamics of the three states where, 
\begin{align}
    \mathcal{P}_{n, \textrm{MMST}} = \frac{1}{2}\left(q_n^2 + p_n^2 - \gamma \right) \text{,} 
\end{align}
for the MMST representation where for $N$ electronic states, $\gamma = 2(r_s-1)/N$ and, 
\begin{align}
    \mathcal{P}_{n, \textrm{SM}} = \frac{1}{N} + r_s \left[ \sum_{m = n+1}^N \left(\sqrt{\frac{2}{m(m-1)}}\Omega_{\gamma_m}\right)  -\sqrt{\frac{2(n-1)}{n}}\Omega_{\gamma_n} \right] \text{,} 
\end{align}
%need to check how ive defined gamma for this 
for the spin-mapping representation where $\boldsymbol{\Omega}$ is defined in Eqn~\eqref{omega}. 

These results agree with the exact results and with previous literature,\cite{bossionNonadiabaticMappingDynamics2022, coronadoUltrafastNonadiabaticDynamics2001} and we conclude that the generalised Spin-MInt algorithm is accurate. The results were calculated using a 1.0 timestep and $10^4$ trajectories. The exact results were calculated using the split-operator method with a 1.0 timestep.\cite{Coronado2000, kosloffTimedependentQuantummechanicalMethods1988} 
The high accuracy of the MInt and Spin-MInt algorithms allows for coarser timesteps to be utilised, as seen in the Supplementary Material Figure S.14, where we use timesteps of $10$ and $100$ which agree with the exact result. The energy conservation is second-order, seen in Figure S.13, which alongside the equality to the MInt shows that the Spin-MInt is symplectic for multiple electronic states.

% For models B/C, $5\times 10^5$ trajectories were required in the literature,\cite{bossionNonadiabaticMappingDynamics2022} so the Spin-MInt algorithm provides faster convergence of these results at this timestep. This is as less trajectories are required to average out errors caused by an inaccurate propagator. The Spin-MInt allows for coarser timesteps to be utilised, as seen in the Supplementary Material Figure S.14, where we use timesteps of $10$ and $100$ which agree with exact result. The energy conservation is second-order, seen in Figure S.13, which alongside the equality to the MInt shows that the Spin-MInt is symplectic for multiple states.

\subsection{Computational Timings} \label{comp-timings}
    In Table S.6 of the Supplementary Material, we present computational timings for the MInt and Spin-MInt algorithms using the two-state system-bath and three-state models. The Spin-MInt algorithm is faster than the MInt in all cases, which motivates utilisation of this algorithm for spin-mapping systems. For one nuclear mode ($F=1$) and two-states, this is an approximate 20\% speed up but including 100 nuclear modes ($F=100$) increases this gap to approximately 50\%. This arises due to the rotations to and from the adiabatic basis for each nuclear derivative, Eqns.~\eqref{p_int}--\eqref{EandF}, and as the electronic variables `sandwich' $\bo{E}^k/\bo{F}^k$ in Eqn.~\eqref{p-mint-final} for the MInt algorithm. However, for the Spin-MInt, $\boldsymbol{\Upsilon}$ [Eqn.~\eqref{upsilon}] is not dependent on a nuclear derivative. Hence, the multiplication with the electronic variables can be precomputed once per timestep before the dot product with $\bo{H}^k$ which drastically reduces the cost in Eqn.~\eqref{P-h2-prop}. For the three-state model, we see an approximate 10\% speed up using the Spin-MInt. The three-state model uses a generalised $N$-state Spin-MInt code that computes the basis rather than the explicit two-state form which we believe contributes to the additional cost alongside the $N^2-1$ scaling of the spin-vector. Again, we would expect this gap to widen when including more nuclear modes. 
    
    % However, there are competing factors at play here as the Spin-MInt has poorer scaling than the MInt in the electronic states, due to diagonalising an $N^2-1 \times N^2-1$ matrix compared to the $N \times N$ matrix for the MMST representation. On the other hand, the MInt scaling with the nuclear modes is worse than the Spin-MInt.  
    
    % As mentioned earlier, fewer trajectories are required to converge the correlation functions when sampling spin-mapping variables and the SM-NRPMD correlation function is closer to the exact answer than NRPMD.\cite{Bossion2021} Hence, for large systems, sampling spin-mapping variables, converting to Cartesian variables to use the MInt algorithm and calculating the spin-mapping observables is likely to be the most cost efficient approach for symplectic propagation. 
    
    In comparison to previous work,\cite{Cook2023} this implementation of the MInt is faster due to the ordering of sub-Hamiltonian propagation in the flow map, Eqn.~\eqref{flow-map-mint}, which has a lower computational cost.\cite{Church2018} In Table S.7, we consider the alternative sub-Hamiltonian ordering to compare with the Spin-SL algorithm. This implementation has been improved by saving the propagation matrices used in the second $H_2$ half-timestep for use in the first $H_2$ half-timestep of the subsequent step such that only one additional calculation of the matrices is required. Numerically, a small increase in the cost is seen for the Spin-MInt and MInt algorithms. The Spin-SL algorithm is faster than the Spin-MInt in Table S.5 for $F=1$, but when $F=100$ this becomes negligible.

    To investigate the scaling, we consider the dominant cost as $F$ nuclear states and $N$ electronic states become large. The MInt scales on the $\mathcal{O}(FN^3)$ due to the rotations to and from the adiabatic basis in the momentum propagation. In contrast, a naive implementation of the Spin-MInt would scale as $\mathcal{O}(N^6 + FN^2)$ due to the diagonalization of $\bo{W}$ and $F$ dot products of $N^2-1$ length vectors. While the results here use this naive implementation, this cost can be reduced by using the diagonalization of $\bo{V}$ to diagonalize $\bo{W}$ due to the relation in Eqn.~\eqref{W-mat-relat}, as shown in the Supplementary Material section IB. This avoids the need to compute $\bo{W}$ and could reduce the diagonalization cost to $\mathcal{O}(N^4)$ such that the Spin-MInt would scale as $\mathcal{O}(N^4 + FN^2)$. The Spin-SL has the same scaling as the Spin-MInt, where the only additional cost in the Spin-MInt is integration of the exponential eigenvalues. For large $F$, this is negligible such that the Spin-MInt is of similar cost to the Spin-SL but is symplectic. We leave a full exploration of this as future work and testing on systems where the potential is calculated on-the-fly may provide further insights. As most realistic systems consider $F \gg N$,\cite{Runeson2020,runeson_exciton_2024,muh_refined_2012, sallerImprovedPopulationOperators2020, Shalashilin2011, Mannouch2022} we believe the Spin-MInt will be computationally faster than the MInt whilst providing accurate and symplectic integration (even if calculating MMST observables). We note that the python code for the Spin-MInt algorithm is available as stated in the Data Availability.

\section{Conclusions} \label{conclusions}
We have presented the first known symplectic algorithm for direct propagation of the spin-vector using the spin-mapping Hamiltonian. This is a significant improvement on the angle-based algorithm which requires small timesteps to accurately propagate the variables, is not symplectic and does not satisfy Liouville's theorem. The Spin-MInt provides the same evolution as the MInt algorithm, where both accurately capture the correlation functions,\cite{Cook2023} satisfy Liouville's theorem, are symplectic and have second-order energy conservation. We have shown that the assumption prior to this work that structure preservation results in symplecticity is not sufficient for the non-trivial coupling between the nuclear canonical and electronic spin-systems. We do this through deriving the Spin-SL algorithm in Appendix~\ref{Spin-SL-algo}, which despite providing accurate trajectories and being structure preserving is not symplectic. Instead, we have rigorously proven symplecticity (Supplementary Material Section IF) by transforming into conjugate canonical variables.

We have generalised the Spin-MInt algorithm to  multiple electronic states using the spin-coherent states from the literature.\cite{bossionNonadiabaticMappingDynamics2022, Runeson2020} For a three-state Morse potential, we accurately capture the population dynamics at larger timesteps.\cite{bossionNonadiabaticMappingDynamics2022}  We find that the Spin-MInt is faster than the MInt algorithm for all models tested here, especially when considering large nuclear DoF. There are competing costs in the Spin-MInt as one includes more electronic and nuclear DoF. However, for realistic systems where there are more nuclear DoF than electronic states, we expect the Spin-MInt to be faster. Future work would include further investigation of scaling with system size. Therefore, the Spin-MInt algorithm should be utilised in future spin-mapping simulations to ensure fast, accurate and symplectic propagation.

% Whilst one would expect the Spin-MInt to be more expensive for $N$ electronic states due to the $N^2-1$ spin-variables compared to the $2N$ for the MInt, we find it to be cheaper for three-states. However, as the Spin-MInt has significantly better scaling with nuclear DoF compared to the MInt, 

\section*{Supplementary Material}
The supplementary material comprises of three sections; additional algebra, model parameters and sampling, and additional results. The additional algebra section includes decomposition of $\bo{V}$ and $\bo{W}$ when $\Delta$ is real, an alternative Spin-MInt nuclear momentum propagation, an explicit form of $e^{-i\bo{W}t}$ and related integral and derivatives when $\Delta$ is real, proof of satisfaction of Liouville's theorem, symplecticity and time-reversibility for the Spin-MInt algorithm. The model section tabulates model parameters and details the sampling. The additional results section consists of further single trajectories (variable propagation and symplecticity at different timesteps), ensemble results (symplecticity, Liouville's theorem, spin-magnitude, energy conservation, correlation functions and populations) and tabulated computational timings. 

\section*{Acknowledgements}
TJHH acknowledges a Royal Society University Research Fellowship URF\textbackslash R1\textbackslash 201502. LEC acknowledges a University College London studentship. JRR acknowledges a summer studentship funded by TJHH's RS URF and funding from the Engineering and Physical Sciences Research Council [grant number EP/Z534882/1]. We thank Pengfei (Frank) Huo and Duncan Bossion for sharing details of the angle-based algorithm.

\section*{Author Declaration}
The authors have no conflicts to disclose.

\section*{Data Availability} \label{data-avaliability}
The data that support the findings of this study are openly available in UCL Research Data Repository at http://doi.org/[to be inserted], reference number [to be inserted].

\appendix
    \renewcommand{\thesubsection}{\Alph{subsection}}
    \section*{Appendices}
    \addcontentsline{toc}{section}{Appendices}
    \numberwithin{equation}{subsection}

    \subsection{MInt Monodromy Matrices}
    \label{Monodromy Matrices}

    For the MMST variable system with one nuclear DoF, the monodromy matrix is,\cite{Cook2023, Church2018}
\begin{gather} \label{monoelementform}
    \bo{M} = \left[ {\begin{array}{cccc}
    \textrm{M}_{RR} & \bo{M}_{R\bo{q}} & \textrm{M}_{RP} & \bo{M}_{R\bo{p}}  \\
    \bo{M}_{\bo{q}R} & \bo{M}_{\bo{qq}} & \bo{M}_{\bo{q}P} & \bo{M}_{\bo{qp}}  \\
    \textrm{M}_{PR} & \bo{M}_{P\bo{q}} & \textrm{M}_{PP} & \bo{M}_{P\bo{p}} \\
    \bo{M}_{\bo{p}R} & \bo{M}_{\bo{pq}} & \bo{M}_{\bo{p}P} & \bo{M}_{\bo{pp}}  \\
  \end{array} } \right] \text{,}
\end{gather}
such that the monodromy matrix for $H_{1, \textrm{MMST}}$ using the MInt is
    \begin{align}
        \label{H1Mono}
        \bo{M}_{\textrm{M}, H_{1, \textrm{MMST}}} = \begin{bmatrix}
        1 & \bo{0}^\mathrm{T} & \Delta t/2m & \bo{0}^\mathrm{T}  \\
        \bo{0} & \mathbb{I} & \bo{0} & \mathbb{O}  \\
        0 & \bo{0}^\mathrm{T} & 1 & \bo{0}^\mathrm{T}  \\
        \bo{0} & \mathbb{O} & 0 & \mathbb{I}  \\
         \end{bmatrix} \text{,}
    \end{align}
where $\bo{0}^\textrm{T} = [0,0]$, the determinant is unity and, satisfies the symplecticity criterion in Eqn.~\eqref{symplecticity-eqn}.\cite{Cook2023, Church2018}

The monodromy matrix for $H_{2, \textrm{MMST}}$ can be found by defining,
\begin{subequations} \label{monoelements} \begin{align}
    \bo{a} &= -\bo{p}^\textrm{T}\bo{E} + \bo{q}^\textrm{T}\bo{F} \text{,} \\
    b &= \begin{aligned}[t] - \Delta t \bo{V}_0{''} {} -\frac{1}{2}\left( \bo{q}^\textrm{T}\bo{E}'\bo{q} + \bo{p}^\textrm{T}\bo{E}'\bo{p} - 2\bo{q}^\textrm{T}\bo{F}'\bo{p}\right)  + {} \frac{1}{2} \mathrm{Tr}[\bo{{V}}'']\Delta t \text{,} \end{aligned} \\
    \bo{e} &= -\bo{q}^\textrm{T}\bo{E} -\bo{p}^\textrm{T}\bo{F} \text{,}\\
    \bo{f} &= \bo{C}'\bo{p} +\bo{D}'\bo{q} \text{,}\\
    \bo{g} &= \bo{C}'\bo{q} - \bo{D}'\bo{p} \text{,}
    \end{align}
\end{subequations}
where prime indicates the derivative with respect to $R$ such that for the MInt,\cite{Church2018, Cook2023} 
\begin{equation}
    \label{H2monomint}
    \bo{M}_{\textrm{M},H_{2, \textrm{MMST}}} = \left[ {\begin{array}{cccc}
    1 & \pmb{0}^\textrm{T}  & 0 & \pmb{0}^\textrm{T}  \\
    \bo{g}  & \bo{C} & \pmb{0} & -\bo{D}  \\
    b & \bo{e} & 1 & \bo{a} \\
    \bo{f} & \bo{D} & \pmb{0} & \bo{C}  \\
  \end{array} } \right]  \text{,}
  % \quad \text{and}\quad |\bo{M}_{H_{2}}| = 1.
 \end{equation}
which is symplectic and satisfies Liouville's theorem,\cite{Church2018, Cook2023} such that the overall propagation of $H_{1, \textrm{MMST}}$ and $H_{2, \textrm{MMST}}$ is symplectic.\cite{Cook2023, Leimkuhler2005}
  
\subsection{Spin-MInt Monodromy Matrices}
\label{symp-spin-mint}

For $H_{1, \textrm{SM}}$, only $R$ is propagated such that $\bo{M}_{\textrm{SM}, H_{1, \textrm{SM}}}$ is the same as for the MInt, Eqn.~\eqref{H1Mono} as a $4\times4$ matrix for one nuclear DoF, and is easily shown to be symplectic.\cite{Cook2023} 

For $H_{2, \textrm{SM}}$, this is significantly more complex. Defining the exponential matrix as, 
\begin{align}
     \bo{Q} = e^{-i\bo{W} \Delta t} = \bo{S}_We^{-i\boldsymbol{\Lambda}_W\Delta t}\bo{S}^\dagger_W 
    \text{,}
\end{align}
such that the $n$th row of $e^{-i\bo{W}t}$ is the vector $\bo{Q}_n$. The propagation of $\bo{u}$ is, 
\begin{align}
    u_x (t) & = \bo{Q}_1 \bo{u} (0) \text{,}\\
    u_y (t) & = \bo{Q}_2 \bo{u} (0) \text{,} \\
    u_z (t) & = \bo{Q}_3 \bo{u} (0) \text{.}
\end{align}

Repeating for the derivative of $e^{-i\bo{W}t}$ with respect to $R$,
\begin{align}
    \bo{Q}' = (e^{-i\bo{W} \Delta t })^{'} &= \bo{S}'_We^{-i\boldsymbol{\Lambda}_W\Delta t}\bo{S}^\dagger_W  -i\Delta t \bo{S}_We^{-i\boldsymbol{\Lambda}_W\Delta t}\boldsymbol{\Lambda}'_{W}\bo{S}^\dagger_W + \bo{S}_We^{-i\boldsymbol{\Lambda}_W\Delta t}\bo{S}'^\dagger_W
\end{align}
such that $\bo{Q}'_n$ is the $n$-th row of $\bo{Q}'$. The eigenvalues, eigenvectors, and their derivatives are presented in Supplementary Material Section IB for the $N=2$ system where $H_y =0$. For $N>2$, there exists algorithms in the literature to compute these.\cite{Church2018, magnusDifferentiatingEigenvaluesEigenvectors1985} We define the derivatives of $\bo{u}$ with respect to the conjugate variables as, 
\begin{subequations}
\begin{align}
    \bo{v}_1 &= \frac{\partial \bo{u}}{\partial r_s \cos\theta} = \begin{bmatrix}
        -2\cos\phi/\tan\theta \\
        -2\sin\phi/\tan\theta \\
        2
    \end{bmatrix} \text{,} \\
    \bo{v}_2 &= \frac{\partial \bo{u}}{\partial \phi} = \begin{bmatrix}
        -u_y\\
        u_x \\
        0
    \end{bmatrix} \text{.}
\end{align}
\end{subequations}

We note that, 
\begin{align}
    r_s \cos \theta (t) = \frac{1}{2} \bo{Q}_3 \bo{u} (0) \text{,}
\end{align}
and, 
\begin{align}
\tan \phi (t)= \frac{\bo{Q}_2 \bo{u}(0)}{\bo{Q}_1 \bo{u}(0)} \text{,}
\end{align}
such that derivatives of $\phi$ are,
\begin{align}
    \frac{\partial \phi (t)}{\partial f(0)} =  \frac{u_x^2(t)}{u_x^2(t) + u_y^2(t)} \frac{\partial \tan \phi (t)}{\partial f (0)} \text{,}
\end{align}
where $f$ is a ghost variable and, 
\begin{align}
\frac{\partial \tan \phi (t)}{\partial \phi} = \sec^2 \phi (t) = \frac{u_x^2(t) + u_y^2(t)} {u_x^2(t)} \text{,}
\end{align}
such that derivatives are obtained using the quotient rule.

We define the monodromy matrix as, 
 \begin{align}\label{spn-mint-mono}
 \bo{M}_{\textrm{SM}, H_{2, \textrm{SM}}} = \begin{bmatrix}
        1 & 0 &0 &0 \\
         a &b &0 & c \\
         d & e & 1& f\\
         g & h& 0 &j
    \end{bmatrix} \text{,}
\end{align}
with, 
\begin{subequations}
\begin{align}
    a&= \frac{u_x (\bo{Q}'_2 \bo{u}) - u_y(\bo{Q}'_1 \bo{u})}{u_x^2 + u_y^2} \text{,}\\
    b&= \frac{u_x (\bo{Q}_2 \bo{v}_2) - u_y(\bo{Q}_1 \bo{v}_2)}{u_x^2 + u_y^2} \text{,}\\
    c&= \frac{u_x (\bo{Q}_2 \bo{v}_1) - u_y(\bo{Q}_1 \bo{v}_1)}{u_x^2 + u_y^2} \text{,}\\
    d &= - \Delta t\left(\bo{V}''_{0} +  \frac{1}{2}\mathrm{Tr}[\bo{V}'']\right) - \frac{1}{2} \bo{H}''\bo{S}_W \boldsymbol{\Upsilon}\bo{S}^\dagger_W \bo{u}  - \frac{1}{2} \bo{H}'(\bo{S}'_W \boldsymbol{\Upsilon}\bo{S}^\dagger_W + \bo{S}_W \boldsymbol{\Upsilon}'\bo{S}^\dagger_W + \bo{S}_W \boldsymbol{\Upsilon}\bo{S}'^\dagger_W)\bo{u} \text{,}\\
    e&= - \frac{1}{2} \bo{H}' \bo{S}_W \boldsymbol{\Upsilon}\bo{S}^\dagger_W \bo{v}_2 \text{,} \\
    f&= - \frac{1}{2} \bo{H}' \bo{S}_W \boldsymbol{\Upsilon}\bo{S}^\dagger_W \bo{v}_1  \text{,}\\
    g&=  \frac{1}{2} \bo{Q}'_3 \bo{u} \text{,}\\
    h&= \frac{1}{2} \bo{Q}_3 \bo{v}_2 \text{,}\\
    j&= \frac{1}{2} \bo{Q}_3 \bo{v}_1 \text{,}
\end{align}
\end{subequations}
where $u_x/ u_y/ u_z$ are at evolved time, $\bo{u}/ \bo{v}_1/ \bo{v}_2$ are at initial time and,
\begin{align}
    \Upsilon'_{nm} = \begin{cases}
         \frac{i \lambda_n'}{\lambda_n^2}\left[-i\Delta t \lambda_n e^{-i\lambda_n \Delta t} -(e^{-i\lambda_n \Delta t}-1)\right] &\quad n=m \quad \textrm{and}\quad \lambda_n \neq 0 \\
         0 &\quad \textrm{otherwise}\text{,}
    \end{cases}
\end{align}
is obtained from differentiation of Eqn.~\eqref{upsilon}. The determinant is,
\begin{align}
    |\bo{M}_{\textrm{SM}, H_{2, \textrm{SM}}}| = jb-ch = 1 \text{,}
\end{align}
as shown in the Supplementary Material Section IE such that the Spin-MInt satisfies Liouville's theorem. The symplecticity criterion is, 
\begin{align} \label{SM-symp-cond}
    \bo{M}^\mathrm{T}_{\textrm{SM}, H_{2, \textrm{SM}}} \bo{J}^{-1} \bo{M}_{\textrm{SM}, H_{2, \textrm{SM}}} 
    & = \begin{bmatrix}
        0 & C_1 & -1 & C_2 \\
        -C_1 & 0 &0 &-1 \\
        1 &0&0&0 \\
        -C_2 & 1 & 0 &0
    \end{bmatrix} \text{,}
\end{align}
where we show that $C_1 = -e-ah+gb = 0$ and $C_2 =-f-ja+cg =0$ in Section IF of the Supplementary Material, such that the Spin-MInt is rigorously symplectic.

    \subsection{Angle-based Algorithm} \label{theta-phi-algo}
    An alternative algorithm for propagating spin-mapping variables is to propagate the angles directly.\cite{Bossion2021,bossionNonadiabaticMappingDynamics2022} The propagation for $H_{1, \textrm{SM}}$ is identical to the MInt and Spin-MInt. For $H_{2, \textrm{SM}}$, a Velocity-Verlet step is performed in the spin-mapping angles followed by nuclear momentum propagation. 
    The electronic EOM are, 
    \begin{subequations}
    \begin{align}
        r_s \cos \theta (\Delta t/2) &= r_s\cos \theta (0) + 2\Delta r_s \sin \theta (0)\sin \phi(0) \Delta t/2 \text{,} \\
        \phi (\Delta t) &= \phi(0) + \left[ (V_1 - V_2) -2\Delta \frac{\cos \phi (0)}{\tan \theta (\Delta t/2)}\right] \Delta t \text{,}\\
        r_s \cos \theta (\Delta t) &= r_s\cos \theta (\Delta t/2) + 2\Delta r_s \sin \theta (\Delta t/2)\sin \phi(\Delta t) \Delta t/2 \text{,}
    \end{align}
    \end{subequations}
    and the nuclear momentum propagation is, 
    \begin{align}
        P_k(\Delta t) = P_k(0) - \left[ V^k_0 + (V_1^k - V_2^k)r_s \cos \theta (\Delta t)\right] \Delta t \text{.}
    \end{align}
    Alternatively, $\theta$ can be propagated as propagating $r_s \cos \theta$ is equivalent to utilising the chain rule.\cite{Bossion2021} The monodromy matrix is quite complex and for one nuclear DoF is 
    \begin{align}
        \bo{M}_{ {\theta-\phi},H_{2, \textrm{SM}}} = \bo{M}_{P}\bo{M}_{r_s\cos \theta (\Delta t)} \bo{M}_{\phi (\Delta t)}\bo{M}_{r_s\cos \theta (\Delta t/2)} \text{,}
    \end{align}
    where we specify the elements that deviate from the identity matrix, 
    \begin{subequations}
    \begin{align}
        &\bo{M}_{P, 31} = -\left[\bo{V}_0''+ (V_1'' - V_2'')r_s\cos\theta(\Delta t) \right]\Delta t \text{,}\\
        &\bo{M}_{P, 34} = (V_1 ' - V_2 ')r_s \sin \theta(\Delta t)\Delta t \text{,}\\
        &\bo{M}_{r_s\cos \theta (\Delta t), 42} = 2\Delta r_s \sin \theta (\Delta t/2) \cos \phi (\Delta t) \Delta t/2 \text{,}\\
        &\bo{M}_{r_s\cos \theta (\Delta t), 44} =  1- 2\Delta \frac{\sin \phi (\Delta t)}{\tan \theta (\Delta t/2)}\Delta t/2 \text{,}\\
        &\bo{M}_{\phi (\Delta t), 21} = (V_1 ' - V_2')\Delta t \text{,}\\
        &\bo{M}_{\phi (\Delta t), 22} =  1 + 2\Delta\frac{\sin \phi (0)}{\tan \theta (\Delta t/2)}\Delta t \text{,}\\
        &\bo{M}_{\phi (\Delta t), 24} = -2\Delta \frac{\cos \phi (0)}{r_s \sin ^3 \theta (\Delta t/2)}\Delta t \text{,}\\
        &\bo{M}_{r_s\cos \theta (\Delta t/2), 42} = 2\Delta r_s \sin \theta (0) \cos \phi (0) \Delta t/2 \text{,} \\
        &\bo{M}_{r_s\cos \theta (\Delta t/2), 44} =  1- 2\Delta \frac{\sin \phi (0)}{\tan \theta (0)}\Delta t/2 \text{,}
    \end{align}
    \end{subequations}
   and the determinant is, 
    \begin{align} \label{theta-phi-det}
        |\bo{M}_{ {\theta-\phi},H_{2, \textrm{SM}}}| &= \bo{M}_{r_s\cos \theta (\Delta t), 44} \times \bo{M}_{\phi (\Delta t), 22} \times \bo{M}_{r_s\cos \theta (\Delta t/2), 44} \\
        &= 1 + \mathcal{O}(\Delta t) \text{,}
    \end{align}
    such that Liouville's theorem is satisfied on the order of $\Delta t$. Evaluating the symplecticity criterion for each of these matrices results in terms that differ from the criterion on order of $\Delta t$. Hence, this algorithm is unlikely to be symplectic for a finite timestep.

\subsection{Split-Liouvillian Spin Algorithm}\label{Spin-SL-algo}
One can derive a Split-Liouvillian (Spin-SL) algorithm for spin-mapping by using a Suzuki-Trotter decomposition. Despite previous claims that this will make the algorithm symplectic,\cite{Richardson2013, Richardson2017, tranchidaMassivelyParallelSymplectic2018} we can algebraically show this does not result in a symplectic algorithm when considering a non-trivial coupling between the electronic and nuclear DoF. Further splitting the Liouvillian for the sub-Hamiltonian $H_2$ does not result in exact sub-Hamiltonian evolution,\cite{Leimkuhler2005} and the energy will drift during simulation.\cite{huangSymplecticSpinLatticeDynamics2025, Cook2023}

We can define the Liouvillian for $H_{2, \textrm{SM}}$ in canonical variables as
\begin{align}
    \mathcal{L}_2 = \sum_{i=1}^F\dot{P_i}\frac{\partial}{\partial P_i} + \dot{\phi}\frac{\partial}{\partial \phi} + \dot{r \cos \theta} \frac{\partial}{\partial r\cos\theta} \textrm{,}
\end{align}
which we can further split into \begin{subequations}
    \begin{align}
        \mathcal{L}_{\bo{P}} 
        &= \sum_{i=1}^F\dot{P_i}\frac{\partial}{\partial P_i} \textrm{,} \\
        \mathcal{L}_\mathrm{el} &= \dot{\phi}\frac{\partial}{\partial \phi} + \dot{r \cos \theta} \frac{\partial}{\partial r\cos\theta} \textrm{,}
    \end{align}
\end{subequations}
such that the evolution is
\begin{align} \label{SLflow}
    \Psi_{H_{\textrm{SM}}, \Delta t}^{\textrm{SL-Spin}} &:= e^{\mathscr{L}_{\textrm{el}} \frac{\Delta t}{2}} e^{\mathscr{L}_{\bo{P}} \frac{\Delta t}{2}} e^{\mathscr{L}_{1} \Delta t} e^{\mathscr{L}_{\bo{P}} \frac{\Delta t}{2}}e^{\mathscr{L}_{\textrm{el}} \frac{\Delta t}{2}} \text{,} %\label{eqn14b}
\end{align}
where a Suzuki-Trotter decomposition is used,\cite{Richardson2013, Cook2023, Church2018, tranchidaMassivelyParallelSymplectic2018} and $\mathscr{L}_{1}$ is the Liouvillian for $H_{1, \textrm{SM}}$ resulting the EOM in Eqn.~\eqref{R-prop-h1} for a whole timestep. The corresponding flowmap is, 
\begin{align} \label{SLflow2}
    \Psi_{H_{\textrm{SL}}, \Delta t}^{\textrm{SL-Spin}} &:= \Psi_{\textrm{SL-Spin}^b, \frac{\Delta t}{2}} \circ \Phi_{H_{1}, \Delta t} \circ \Psi_{\textrm{SL-Spin}^a, \frac{\Delta t}{2}} \text{,}
\end{align}
where $\Psi_{\textrm{SL-Spin}^{a/b}}$ represents the approximate propagation of $H_{2, \textrm{SM}}$. We have used a symmetric splitting such that the two half timesteps are not equal and the superscripts $a$ is $e^{\mathscr{L}_{\bo{P}}}  e^{\mathscr{L}_{\textrm{el}}}$ and $b$ is $e^{\mathscr{L}_{\textrm{el}}}  e^{\mathscr{L}_{\bo{P}}}$. The EOM for $H_{2, \textrm{SM}}$ are then,
\begin{subequations}
\begin{align}
    \dot{P_k} &= -V^k_0 - \frac{1}{2}(\mathrm{Tr}[\bo{V}^k]+\bo{H}^k\bo{u}) \textrm{,}\\
    \dot{r\cos\theta} &=\frac{1}{2}(H_xu_y - H_yu_x) \textrm{,}\\
    \dot{\phi} &=H_z -H_x \frac{\cos\phi}{\tan \theta} - H_y \frac{\sin\phi}{\tan \theta} \textrm{,}
\end{align}
\end{subequations}
where the electronic EOM are easily shown to be equivalent to Eqn.~\eqref{Heisenberg},\cite{Bossion2021, bossionNonadiabaticMappingDynamics2022} such that integration is Eqn.~\eqref{u-prop}, for half a timestep. As this is identical to the Spin-MInt, this algorithm will be structure preserving but not necessarily symplectic. 

The nuclear momentum can be integrated as
\begin{align}
    P_k(\Delta t/2) = P_k(0) - \frac{\Delta t}{2} \left\{V^k_0 + \frac{1}{2}\left[\mathrm{Tr}[\bo{V}^k]+\bo{H}^k\bo{u}(\Delta t/2)\right]\right\} \textrm{,}
\end{align}
where the electronic variables are the half timestep evolved values.

We now test the propagation of $H_{2, \textrm{SM}}$ for symplecticity and for simplicity, we use one nuclear DoF. For the electronic evolution 
\begin{align}
 \bo{M}_{\textrm{el}} = \begin{bmatrix}
        1 & 0 &0 &0 \\
         a &b &0 & c \\
         0 & 0 & 1& 0\\
         g & h& 0 &j
    \end{bmatrix} \textrm{,}
\end{align}
where the elements are the same as for the Spin-MInt where $\bo{u}$ is half timestep evolved and $|\bo{M}_{\textrm{el}}| =1$ as shown for the Spin-MInt. Evaluating the symplecticity results in
\begin{align}
    \bo{M}_{\mathrm{el}}^\mathrm{T} \bo{J}^{-1} \bo{M}_{\mathrm{el}} = \begin{bmatrix}
        0 & bg-ah &-1 &cg-aj \\
         -bg+ah &0 &0 & -1 \\
         1 & 0 & 0& 0\\
         -cg+ja & 1& 0 &0
    \end{bmatrix} \textrm{,}
\end{align}
which will be symplectic if $bg-ah = 0$ and $cg-ja =0$. As Eqn~\eqref{SM-symp-cond} is satisfied, the above will not be symplectic for a nuclear-dependent potential. 

For the nuclear momentum 
\begin{align}
 \bo{M}_{{P}} = \begin{bmatrix}
        1 & 0 &0 &0 \\
         0 &1 &0 & 0 \\
         \tilde{d} & \tilde{e} & 1& \tilde{f}\\
         0 & 0& 0 &1
    \end{bmatrix} \textrm{,}
\end{align}
where, 
\begin{subequations}
\begin{align}
    \tilde{d} &= -\frac{\Delta t}{2}\left\{\bo{V}_0'' +\frac{1}{2}\left[\mathrm{Tr}[\bo{V}''] + \bo{H}''\bo{u}(\Delta t/2)\right]\right\}\textrm{,}\\
    \tilde{e} &= - \frac{\Delta t}{4} \bo{H}' \bo{v}_2(\Delta t/2) \textrm{,}\\
    \tilde{f} &= - \frac{\Delta t}{4} \bo{H}' \bo{v}_1(\Delta t/2) \textrm{,}
 \end{align}   
\end{subequations}
and it is easy to see $|\bo{M}_{\textrm{P}}| =1$. Evaluating symplecticity results in 
\begin{align}
    \bo{M}_{P}^\mathrm{T} \bo{J}^{-1} \bo{M}_{P} = \begin{bmatrix}
        0 & -\tilde{e} &-1 & -\tilde{f} \\
         \tilde{e} &0 &0 & -1 \\
         1 & 0 & 0& 0\\
         \tilde{f} & 1& 0 &0
    \end{bmatrix} \textrm{,}
\end{align}
which will be symplectic if $\tilde{e} = \tilde{f}=0$ which will only be satisfied if there is no nuclear dependence of the potential matrix. 

The combination of the two matrices is 
\begin{align}
 \bo{M}_{\mathrm{el}}\bo{M}_{{P}} = \begin{bmatrix}
        1 & 0 &0 &0 \\
         a &b &0 & c \\
         \tilde{d} & \tilde{e} & 1& \tilde{f}\\
         g & h& 0 &j
    \end{bmatrix} \textrm{,}
\end{align}
which compared to Eqn.~\eqref{spn-mint-mono} will only be symplectic if $e = \tilde{e}$ and $f = \tilde{f}$ which requires all eigenvalues of $\bo{W}$ being $0$ or that the potential matrix is independent of nuclear position. 

The SL-Spin algorithm will not be symplectic as $\bo{M}_\textrm{el}$, $\bo{M}_P$ and the combination of $\bo{M}_{\mathrm{el}}\bo{M}_{{P}}$ are not symplectic. However, it will satisfy phase-space preservation through Liouville's theorem and due to the same electronic propagation as the Spin-MInt will preserve the Poisson structure. Hence, structure preservation will not automatically lead to a symplectic algorithm when the Hamiltonian is coupled such that splitting the Liouvillian results in a loss of exact sub-Hamiltonian evolution.

\subsection[Generalisation to N Electronic States]{Generalisation to \textit{N} Electronic States} \label{multiple spin mint}

Here, we generalise the Spin-MInt algorithm to multiple electronic states, $N$, by shifting the formalism from the 3 dimensional Lie group SU(2) to the $N^2-1$ dimensional SU($N$).\cite{bossionNonadiabaticMappingDynamics2022, Runeson2020}

We define a basis of the Lie algebra $\mathfrak{su}(N)$ which replaces the Pauli spin matrices. These are the Generalised Gell-Mann (GGM) matrices,\cite{bossionNonadiabaticMappingDynamics2022, gell-mannSymmetriesBaryonsMesons1962,tilmaSUNsymmetricQuasiprobabilityDistribution2011} where we obtain $N^2 -1$ matrices, $\hat{S}_i$ where $i \in \{1, \hdots,N^2 -1 \}$. This comprises of $N(N-1)/2$ symmetric matrices, 
\begin{align}
    \hat{S}_{\alpha_{nm}} = E_{mn} + E_{nm} \text{,}
\end{align}
$N(N-1)/2$ antisymmetric matrices, 
\begin{align}
    \hat{S}_{\beta_{nm}} =-i( E_{mn} - E_{nm}) \text{,}
\end{align}
and $N-1$ diagonal matrices, 
\begin{align}
    \hat{S}_{\gamma_{n}} = \sum_{l=1}^{n-l} \left(  \sqrt{\frac{2}{n(n-1)}} E_{ll}\right) - \sqrt{\frac{2(n-1)}{n}} E_{nn} \text{,}
\end{align}
where $E_{nm}$ is a $N \times N$ matrix with 1 in the $(n,m)$-th position and 0 elsewhere and,
\begin{subequations}
\begin{align}
    \alpha_{nm} &= n^2 + 2(m-n)-1 \text{,}\\
    \beta_{nm} &= n^2 + 2(m-n) \text{,}\\
    \gamma_{n} &= n^2 -1 \text{,}
\end{align}
\end{subequations}
for $1 \leq m < n \leq N$ and $2 \leq n \leq N$.\cite{bossionNonadiabaticMappingDynamics2022} These are ordered conventionally following the literature.\cite{bertlmannBlochVectorsQudits2008, pfeiferLieAlgebrasSuN2003a}

The spin-vector for $N$ states as spin-coherent states is,\cite{bossionNonadiabaticMappingDynamics2022, Runeson2020}
\begin{align}
    \ket{\boldsymbol{\Omega}} = \sum_{n=1}^N  \ket{n} \braket{n}{\boldsymbol{\Omega}} \text{,}
\end{align}
where the expansion coefficients are, 
\begin{align}
    \braket{n}{\boldsymbol{\Omega}}_N = \begin{cases}
        \braket{n}{\boldsymbol{\Omega}}_{N-1}    & 1 \leq n < N-1 \\
        \braket{N-1}{\boldsymbol{\Omega}}_{N-1} \cos \frac{\theta_N}{2}  & n = N-1\\
        \braket{N-1}{\boldsymbol{\Omega}}_{N-1} e^{i \phi_N} \sin \frac{\theta_N}{2} & n = N\text{,}
    \end{cases}
\end{align}
which are defined recursively from $\braket{1}{\boldsymbol{\Omega}}_1 = 1$ such that, 
\begin{align} \label{omega}
    \Omega_{\alpha_{nm}} = \bra{\boldsymbol{\Omega}} \hat{S}_{\alpha_{nm}} \ket{\boldsymbol{\Omega}}\text{,}
\end{align}
and likewise for the antisymmetric and diagonal GGM matrices.\cite{Runeson2020} The angles are the general Euler angles in a multidimensional Bloch sphere where $N-1$ spheres are obtained such that $\theta_n \in [0, \pi]$ and $\phi_n \in [0, 2\pi]$ and further details can be found in the literature.\cite{Runeson2020, bossionNonadiabaticMappingDynamics2022} 

Here, we express the $N^2-1$ dimensional spin-vector in terms of the MMST variables where, 
\begin{subequations}
\begin{align}\label{MMSTToSpinGeneral}
    2 r_s \Omega_{\alpha_{nm}} &= (\bo{q} - i\bo{p})^\mathrm{T} \hat{S}_{\alpha_{nm}} (\bo{q + i\bo{p}}) \\
    &= \mathrm{Tr}[\hat{S}_{\alpha_{nm}} \bo{C}(\bo{q}, \bo{p})] \text{,}
\end{align}
\end{subequations}
and likewise for the antisymmetric and diagonal GGM matrices where $\bo{C}  = ({\bf{q}} +i{\bf{p}})\bigotimes ({\bf{q}} -i{\bf{p}})^\textrm{T}$.\cite{bossionNonadiabaticMappingDynamics2022} The squared radius of the hyper-sphere is, 
\begin{align}
    R^2 = \sum_{n=1}^N (q_n^2 + p_n^2) \equiv 2 r_s \text{,}
\end{align}
where the radius is often set by the Stratonovich-Weyl kernel used and $r_s$ here is twice what is used earlier for $N=2$.\cite{bossionNonadiabaticMappingDynamics2022, Runeson2019, Runeson2020} For example, with the $W$ kernel used for the three-state model in Figure~\ref{fig:morse}, $r_s = \sqrt{N+1}$ such that $R^2 = 2\sqrt{N+1}$.\cite{Runeson2020, bossionNonadiabaticMappingDynamics2022, Runeson2019} 

The generalised spin-mapping Hamiltonian is, 
\begin{align}
    H_{\textrm{SM}} (\boldsymbol{\Omega}) = \frac{1}{2}\bo{P}^\mathrm{T}\bo{\mu}^{-1}\bo{P} + V_0 + \frac{1}{N}\mathrm{Tr}[\bo{V}] + \frac{1}{2} \bo{H} \cdot (2r_s \boldsymbol{\Omega}) \text{,}
\end{align}
where the diabatic potential energy matrix is $N \times N$ and $\bo{H}$ is an $N^2-1$ dimensional vector.\cite{ bossionNonadiabaticMappingDynamics2022} This is equivalent to the MMST Hamiltonian where $\gamma = 2(r_s-1)/N$. 
$\bo{H}$ can be calculated by, 
\begin{align}
    H_{\alpha_{nm}} = 2\mathrm{Tr}[\bo{V}\hat{S}_{\alpha_{nm}}] \text{,}
\end{align}
and likewise for the antisymmetric and diagonal matrices.\cite{bossionNonadiabaticMappingDynamics2022} Splitting this Hamiltonian results in, 
\begin{subequations}
    \begin{align}
    H_{1, \textrm{SM}} &= \frac{1}{2}\bo{P}^\mathrm{T}\bo{\mu}^{-1}\bo{P} \text{,}\\
    H_{2, \textrm{SM}} &= V_0 + \frac{1}{N}\mathrm{Tr}[\bo{V}] + \frac{1}{2} \bo{H} \cdot (2r_s \boldsymbol{\Omega}) \text{,}
\end{align}
\end{subequations}
such that the propagation of $H_{1, \textrm{SM}}$ for half a timestep is Eqn.~\eqref{R-prop-h1}.
For $H_{2, \textrm{SM}}$, the derivative equations of motion for $\boldsymbol{\Omega}$ can we written in terms of structure constants detailed in Appendix A of Ref.~[\!\!\citenum{bossionNonadiabaticMappingDynamics2022}],
\begin{align} \label{omegadot}
     \dot{\Omega}_i =\frac{-i}{2} \text{Tr}\{ \hat{S}_i [V,\hat{S}_k] \} \Omega_k = f_{ijk} H_j \Omega_k = -i\bo{W}_{ik} \Omega_k \text{.}
 \end{align}
 
The $\bo{W}$ matrix is the adjoint representation of the potential matrix, $\bo{V}$, in the $\mathfrak{su}(N)$ Lie algebra,\cite{pfeiferLieAlgebrasSuN2003a} which can be calculated by, 
\begin{align} \label{W-mat-relat}
    W_{ij} = \frac{1}{2}\mathrm{Tr}\{\hat{S}_i [\bo{V},\hat{S}_j ]\} = i f_{ikj}H_k \text{,}
\end{align}
which is easily shown to give the $\bo{W}$ defined earlier for $N=2$. The integration of Eqn.~\eqref{omegadot} follows Eqn.~\eqref{u-prop}, 
\begin{align}
    \label{omega-prop-eigh}
    \boldsymbol{\Omega}(t+\Delta t) = \bo{S}_We^{-i\boldsymbol{\Lambda}_W\Delta t}\bo{S}^\dagger_W\boldsymbol{\Omega}(t) \text{,}
\end{align}
and the propagation of $\bo{P}$ is, 
\begin{align} 
    P_k(t + \Delta t) &= P_k(t) - \Delta t\left(\bo{V}^k_{0} +  \frac{1}{N}\mathrm{Tr}[\bo{V}^k]\right) - \frac{1}{2} \bo{H}^k \bo{S}_W \boldsymbol{\Upsilon}\bo{S}^\dagger_W\boldsymbol{\Omega}(t) \text{.}
\end{align}
$\boldsymbol{\Upsilon}$ is calculated following Eqn~\eqref{upsilon}, where there are $N^2-1$ eigenvalues as discussed in the Supplementary Material Section IB. The eigenvalue and eigenvector derivatives are not required for trajectory propagation, only when calculating the monodromy, and for $N>2$ algorithms exist in the literature.\cite{magnusDifferentiatingEigenvaluesEigenvectors1985}

The equivalence to the MInt for a general number of states is shown using the GGM basis and taking the derivative of Eqn.~\eqref{MMSTToSpinGeneral},
\begin{subequations}
    \begin{align}
    2r_s\dot{\Omega}_i &= i (\bo{q}-i\bo{p})^\mathrm{T} [\bo{V},\hat{S}_i] (\bo{q}+i\bo{p}) \\
    &=i (\bo{q}-i\bo{p})^\mathrm{T} H_j[\hat{S}_j,\hat{S}_i] (\bo{q}+i\bo{p}) \\
    % &=i (\bo{q}-i\bo{p})^\mathrm{T} iH_jf_{jik}\hat{S}_k (\bo{q}+i\bo{p}) \\
    &= 2r_sf_{ijk}H_j\Omega_k
\end{align}
\end{subequations}
where the total antisymmetry of the structure constants is used. The equivalence of the $\bo{P}$ propagation follows Section~\ref{mint-eqiuvalence} using the GGM basis to construct $\bo{C}(\boldsymbol{\Omega})$. Due to this equivalence for $N$ states, the generalised Spin-MInt will be symplectic for $N>2$.

\bibliography{rp}

\end{document}